\documentclass[draftcls,onecolumn,12pt]{IEEEtran}
\usepackage{color}
\usepackage{graphicx}
\usepackage{epstopdf}
\usepackage{amsmath}
\usepackage{amssymb}
\usepackage[english]{babel}
\usepackage{cite}
\usepackage{rotfloat}
\usepackage{mathtools} 
\usepackage[justification=centering]{caption}
\usepackage{breqn}
\usepackage{amsmath}
\usepackage{etoolbox}
\let\mybibitem\bibitem
\renewcommand{\bibitem}[1]{%
	\ifstrequal{#1}{nature}
	{\color{blue}\mybibitem{#1}}
	{\color{black}\mybibitem{#1}}%
}

\graphicspath{ {Figures/} }

\newtheorem{definition}{Definition}
\newtheorem{theorem}{Theorem}
\newtheorem{remark}{Remark}

\newtheorem{corollary}{Corollary}
\newtheorem{proof}{Proof}

\newcommand{\dis}{\hspace{0.1cm}}
\DeclareCaptionLabelSeparator{periodspace}{.\quad}

\addto\captionsenglish{}
\newcommand\numberthis{\addtocounter{equation}{1}\tag{\theequation}}

\newcommand{\cdf}{\mathbf{\textit{F}}} 
\newcommand{\pdf}{\mathbf{\textit{f}}} 

\newcommand{\hv}{\textbf{\textit{h}}} 
\newcommand{\h}{\textbf{\textit{H}}} 
 
\newcommand{\s}{\textbf{\textit{s}}} 
\newcommand{\p}{\mathbf{\Phi}} 
\newcommand{\prb}{\mathbf{P}}

\newcommand{\z}{\textbf{\textit{z}}} 
\newcommand{\set}{\mathbb{C}} 
\newcommand{\setR}{\mathbb{R}} 
\newcommand{\bp}{\pmb{\phi}} 

\newcommand{\Pdl}{\mathcal{P}^{(T)}} 
\newcommand{\Pdlx}{\mathcal{P}^{(x)}}

\newcommand{\Xm}{X_{max}} 
\newcommand{\Am}{A_{max}} 
\newcommand{\Bm}{B_{min}} 

\begin{document}
\title{Cell Coverage Extension with Orthogonal 
Random Precoding for Massive MIMO Systems}
\author{Nhan~Thanh Nguyen~
        and~Kyungchun~Lee,~\IEEEmembership{Senior Member,~IEEE}
        \thanks{This research was supported by Basic Science Research Program through the National Research Foundation of Korea (NRF) funded by the Ministry of Science, ICT, and Future Planning (NRF-2014R1A1A1002653).}
		\thanks{The authors are with the Department
		of Electrical and Information Engineering, Seoul National University of Science and Technology,
		232 Gongneung-ro, Nowon-gu, Seoul, 01811, Republic of Korea (e-mail:
		nhan.nguyen, kclee@seoultech.ac.kr, corresponding author: Kyungchun Lee).}
}

\maketitle
\begin{abstract}

In this paper, we investigate a coverage extension scheme based on orthogonal random precoding (ORP) for the downlink of massive multiple-input multiple-output (MIMO) systems. In this scheme, a precoding matrix consisting of orthogonal vectors is employed at the transmitter to enhance the maximum signal-to-interference-plus-noise ratio (SINR) of the user. To analyze and optimize the ORP scheme in terms of cell coverage, we derive the analytical expressions of the downlink coverage probability for two receiver structures, namely, the single-antenna (SA) receiver and multiple-antenna receiver with antenna selection (AS). The simulation results show that the analytical expressions accurately capture the coverage behaviors of the systems employing the ORP scheme. It is also shown that the optimal coverage performance is achieved when a single precoding vector is used under the condition that the threshold of the signal-to-noise ratio of the coverage is greater than one. The performance of the ORP scheme is further analyzed when different random precoder groups are utilized over multiple time slots to exploit precoding diversity. The numerical results show that the proposed ORP scheme over multiple time slots provides a substantial coverage gain over the space-time coding scheme despite its low feedback overhead.

\end{abstract}

\begin{IEEEkeywords}
Cell coverage, random beamforming, MIMO, massive MIMO.
\end{IEEEkeywords}
\IEEEpeerreviewmaketitle

\section{Introduction}
In mobile communication, a massive multiple-input multiple-output (MIMO) system, where the base station (BS) is equipped with a large number of antennas, has been recently considered as a potential technique for dramatically improving system performance in terms of spectral and power efficiency \cite{EnergyNgo}, \cite{NoncooperativeMarzetta}. It is also thought that a massive MIMO system is capable of extending its cell coverage by exploiting a large array gain to compensate for the significant path loss in millimeter-wave propagation channels, which provides a wider bandwidth for 5G communication systems \cite{swindlehurst2014millimeter}. Specifically, in the downlink, precoding techniques can be exploited to extend the cell coverage in massive MIMO systems \cite{su2013limited}, \cite{lim2015performance}. 

{Most studies on precoding techniques for MIMO systems have been carried out under the assumption of perfect channel state information (CSI) at the transmitter \cite{collin2004optimal, sampath2001generalized, chiu2010precoding, sidiropoulos2006transmit, chae2009optimality}. However, in practical systems, the CSI is imperfect \cite{love2005limited, love2003grassmannian, choi2014downlink}, and in frequency-division duplexing, it is typically acquired by the feedback signals from the receivers, which results in a significant overhead, especially in massive MIMO systems \cite{lu2014overview, gao2015spatially,choi2015trellis,rao2014distributed}. Moreover, in contrast to unicast data channels, for multicast/broadcast channels, which must be received by a large number of mobile users in each cell,  CSI-based precoding strategies can lead to the potentially excessive overhead \cite{huang2009performance,jindal2006mimo, OnSharif}. Therefore, to achieve the coverage gain in the downlink of massive MIMO systems, non- or partial-CSI based transmission techniques such as random precoding should be considered \cite{obara2014joint}, \cite{nam2012joint}.}

There has been a line of research studying the coverage extension problem. In \cite{CoverageLee}, the authors showed that the cell coverage can be extended by the dual-hop space-time relaying scheme. The results in \cite{EnhancingChen} indicate that the proposed strategy called the  strongest-weakest-normalized-subchannel-first scheduling can significantly expand the coverage of MIMO systems. In \cite{DownlinkDhillon}, the downlink coverage performance in MIMO heterogeneous cellular networks was investigated; furthermore, the work was extended with flexible cell selection in \cite{DownlinkGupta}. The same problem has also been recently considered in massive MIMO systems \cite{AsymptoticBai}, \cite{CellJin}. The analytical expressions for the asymptotic coverage probability and rate for both downlink and uplink in random cellular networks with Poisson distributed BS locations are presented in \cite{AsymptoticBai}. The cell coverage optimization problem for the massive MIMO uplink was investigated in \cite{CellJin}. 

There has been another line of work studying random beamforming. {In \cite{huang2009performance}, the authors presented asymptotic throughput scaling laws for space-division multiple access with orthogonal beamforming known as per user unitary and rate control for the interference- and noise-limited regimes. The work of \cite{OnSharif} showed that in the orthogonal random precoding (ORP) scheme, the throughput scales linearly with the number of transmit antennas $N_t$, provided $N_t$ does not increase faster than $\log n$, where $n$ is the number of users.} The works of \cite{EffectHieu} and \cite{MultiNguyen} investigated the achievable rates in a multi-cell setup subject to inter-cell interference and characterized the achievable degree of freedom region in the MIMO random beamforming scheme. In \cite{OpportunisticViswanath}, the authors proposed the use of multiple transmit antennas with the aim of inducing channel fluctuations to exploit multiuser diversity.

In contrast to the above mentioned approaches, this paper focuses on an ORP scheme to enhance the cell coverage in the downlink of massive MIMO systems. As an advantage, this scheme requires only partial CSI at the transmitter. Specifically, each receiver only feeds back its maximum signal-to-interference-plus-noise ratio (SINR) and the corresponding beam index. The analytical expressions for the downlink coverage probability of the ORP scheme are derived. In addition, the ORP schemes in conjunction with various receive structures and multiple precoder groups are evaluated. We also compare the ORP scheme to the conventional space-time coding (STC) scheme to show that when multiple precoder groups are employed to increase the diversity gain in the ORP scheme, its achievable cell coverage surpasses that of the STC scheme. The analysis results can also be applied to the ORP scheme for multicast/broadcast channels where the CSI cannot be used for precoding.

The outline of the paper is as follows. In Section \ref{sec:Problem Formulation}, we describe the system model and the ORP scheme. Section \ref{sec:Downlink Coverage Probability with ORP} presents an analysis of the downlink coverage probability of the ORP scheme for the cases of single and multiple receive antennas. In Section \ref{Delay Constraint}, the improvement in cell coverage  with multiple random precoder groups under a delay constraint is presented. A comparison between the ORP scheme and STC technique is also given. Section \ref{sec:Numerical Results} provides our simulation results, while Section \ref{Conclusion} concludes the paper.

\textit{Notations}: Throughout this paper, scalars, vectors, and matrices are denoted by lower-case, bold-face lower-case, and bold-face upper-case letters, respectively. The $(i,j)$th element of a matrix is denoted as $\left[\cdot\right]_{i,j}$, and $(\cdot)^T$ and $(\cdot)^*$ denote the transpose and conjugate transpose operators, respectively. Further, $\Vert\cdot\Vert$ denotes the norm of a vector and $\mathbf{E}\left\{\cdot\right\}$ denotes statistical expectation. The distribution of a circularly symmetric complex Gaussian random variable with zero-mean and variance $\sigma^2$ is denoted by $\mathcal{CN}(0,\sigma^2)$, while $\chi^{\eta}\left(\epsilon^2\right)$ denotes the central chi-square random variable of $\eta$ degrees of freedom with mean $\epsilon^2$.  Finally, $\set ^{x \times y}$ and $\setR_{+} ^{2}$ denote the space of $x \times y$ complex matrices and the non-negative real coordinate space of two dimensions, respectively.

\section{Problem Formulation}
\label{sec:Problem Formulation}
\subsection{System Model}
\begin{figure}
	\centering
	\includegraphics[scale = 0.5]{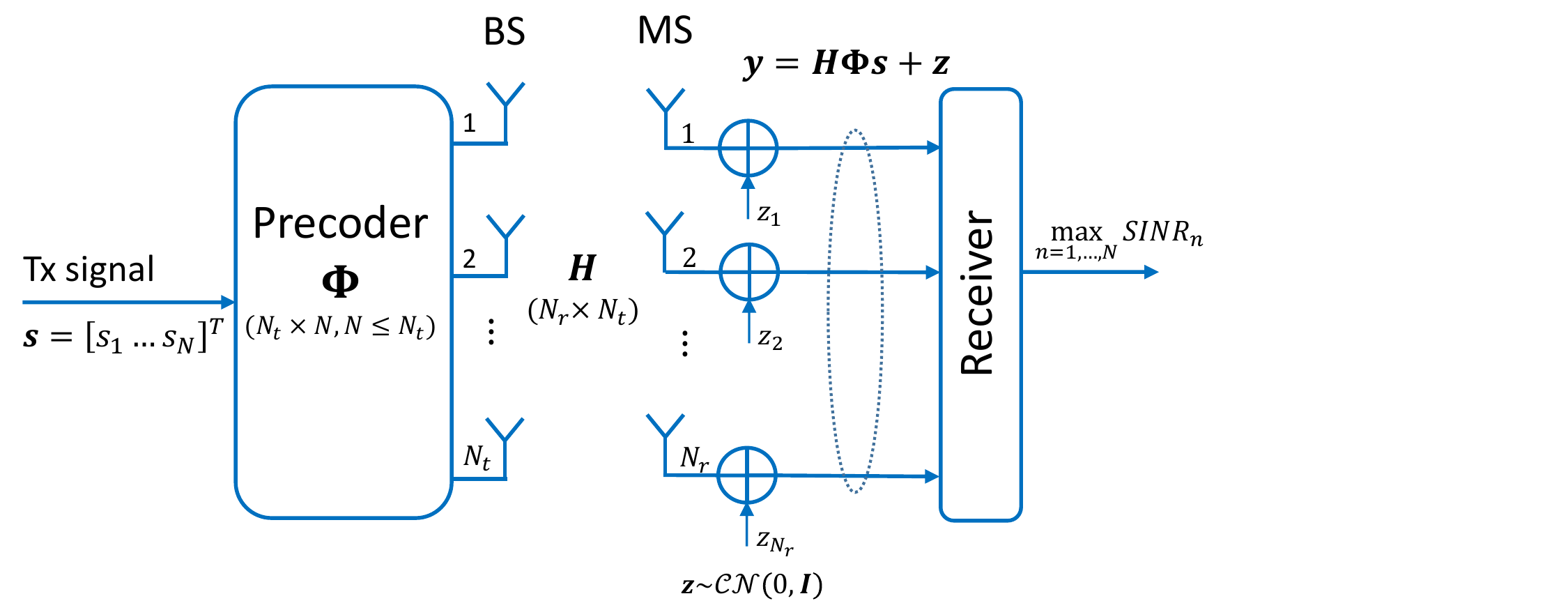}
	\caption{Downlink system with orthogonal random precoding.}
	\label{fig:system_model}
\end{figure}

In this section, we present the system model of a downlink channel in a massive MIMO network, which is illustrated in Fig. \ref{fig:system_model}. The BS and each mobile station (MS) have $N_t$ and $N_r$ antennas, respectively. We assume that the channel is block-fading and is constant during a coherence interval.

At time $t$, the received signal at an MS is given by
\begin{align}
\label{y_rx}
\textbf{\textit{y}}(t) =  \h \p(t) \s(t) + \z(t),
\end{align}
where $\h \in \set^{N_r \times N_t}$ is the channel coefficient matrix between the BS and MS, $\s(t) \in \set^{N \times 1}$ is the vector of transmit symbols, $\z(t) \in \set^{N_r \times 1}$ is an additive white Gaussian noise vector with elements of $\mathcal{CN}(0,\sigma^2)$, and $\p(t) \in \set^{N_t \times N}$ is a random unitary matrix consisting of $N$ orthonormal precoding vectors with the constraint $N \leq N_t$. We assume that the $N$ elements in $\s(t)$ are the signals sent to $N$ different MSs, which implies that $N$ MSs are simultaneously served each time. The channel matrix is assumed to be Rayleigh fading; hence, the coefficients of $\h$ are independent and identically distributed (i.i.d.) Gaussian random variables, i.e., $\left[\h \right]_{i,j} \sim \mathcal{CN}(0,1)$.  Moreover, we assume the average total transmit power is $P_T$, i.e., $\mathbf{E}\left\{\s(t)^* \s(t) \right\} = P_T$, which yields that the transmit power per symbol is $P_T/N$, i.e.,  $\mathbf{E} \left\{ \left|s_i(t)\right|^2 \right\} = P_T/N$, where $s_i(t)$ is the $i$th element in $\s(t)$, $i=1,2,\ldots,N$. Let $\rho$ be the average received signal-to-noise ratio (SNR); then, $\rho$ is expressed as
\begin{align*}
\rho = \frac{\mathbf{E}\left\{\Vert \p(t) \s(t) \Vert^2 \right\}}{\sigma^2} = \frac{P_T}{\sigma^2}.
\end{align*}
{For simplicity, we ignore the inter-cell interference in the signal mode. However, the effect of inter-cell interference can be approximately considered to be included in noise $\textbf{\textit{z}}$. Therefore, by analyzing the coverage for various SNRs, the effect of different inter-cell interference can also be evaluated. Furthermore, depending on the cell deployment, cell-edge users can have different signal strengths, which result in different SNR values. In the subsequent section, to evaluate the coverage performance of the ORP scheme, we analyze the distribution of SINR for various SNR values. Then, the coverage performance is analyzed based on the coverage probability.}

\subsection{Orthogonal Random Precoding}
In the ORP scheme for unicast data channels, the signals are precoded by $N$ orthonormal random precoding vectors before transmission. We describe the ORP scheme by dividing it into two phases as follows:
\subsubsection{Training phase}
\paragraph{Beamforming} The BS randomly generates a precoding matrix $\p$ of $N$ orthonormal precoding vectors $\bp_1$, $\bp_2$, $\ldots$, $\bp_N$. The training signals are multiplied by the precoding matrix before being sent to the MSs.
\paragraph{Beam selection} Each MS computes the SINR of each precoding vector and finds the maximum one. Specifically, at an MS, $N$ SINR values $SINR_{1}$, $\ldots$, $SINR_{N}$, which correspond to $N$ orthogonal beams, are estimated as in the schemes of \cite{EffectHieu}, \cite{OnSharif}. The maximum SINR $SINR_{n}$ is determined, and then the value $SINR_{n}$ as well as its index $n$ is fed back to the BS. In this work, this optimally selected precoding vector is referred to as the ``effective beam."

\subsubsection{Transmission phase}~

When the training phase is finished, the BS knows the effective beam index and its SINR for each MS. Then, if the maximum SINR is higher than a predefined threshold $T$, the MS is determined to be in coverage and can be selected for data transmission. In this phase, the transmit signal is precoded by the effective beam before transmission.

\newpage 
Similar to the unicast data channels, the ORP scheme can also be employed for multicast/broadcast channels in which it is problematic to achieve a coverage gain through multiple transmit antennas because CSI-based precoding schemes cannot be applied. {As an example, in LTE/LTE-A systems, the physical broadcast channel (PBCH), which delivers the master information block to the MSs during the initial call setup procedure \cite{etsi36211, etsi36331}, can be transmitted via random precoding vectors; however, in contrast to the unicast data channels, the training phase is not performed.}

In this work, to analyze the cell coverage extension resulting from the ORP scheme, we use the SINR as a criterion. In other words, it is said that the MS is covered by the BS in the cell if the maximum of $N$ SINRs is larger than threshold $T$. In the next section, we analyze the probability that the user is covered by the BS.
\section{Coverage Probability with the ORP Scheme}
\label{sec:Downlink Coverage Probability with ORP}
\subsection{Single Receive Antenna System}
We first consider the baseline scenario where each MS is equipped with a single receive antenna. Without loss of generality, hereafter, we drop the time index $t$. The received signal in \eqref{y_rx} becomes
\begin{align*}
\textit{y} 
=  \hv^T \p \s + \textit{z}
= \hv^T \sum_{i=1}^{N} \bp_i s_i + \textit{z} \numberthis \label{y_drop_t},
\end{align*}
where $\hv^T \in \set^{1\times N_t}$ is the channel vector between the BS and single-antenna MS, and we assume that the MS estimates $\hv^T \bp_i, i = 1, \ldots, N$, by training procedures. Therefore, each MS can compute the SINR corresponding to the $n$th beam $\bp_n$ by assuming that $s_n$ is the desired signal, while the others are interference from $N-1$ ineffective beams $\bp_i, i \neq n, i = 1, \ldots, N$ \cite{OnSharif}. Specifically, the SINR for $\bp_n$ can be expressed as
\begin{align*}
SINR_{n} 
&=  \frac{\left|  \hv^T \bp_{n}\right|^2 \frac{P_T}{N}}{ \sigma^2 +  \sum_{i \neq n}^N \left| \hv^T \bp_{i}\right| ^2 \frac{P_T}{N}}\\
&= \frac{\left|  \hv^T \bp_{n}\right|^2 }{\frac{N}{\rho} +  \sum_{i \neq n}^N \left| \hv^T \bp_{i}\right| ^2},\hspace{0.1cm} n=1,\ldots, N \numberthis \label{SINR_eqn}.
\end{align*}

The downlink coverage probability can then be defined as follows:
\begin{definition}[Downlink Coverage Probability] 
	\label{def:Pdl}
	In the downlink of a massive MIMO system using the ORP scheme, an MS is said to be in coverage if its maximum  SINR is higher than a predefined threshold $T$. The coverage probability is defined as
	\begin{align}
	\label{P_DL_def}
	\Pdl = \prb \left\{ \max\limits_{n=1,\ldots,N} SINR_n > T \right\}.
	\end{align}
\end{definition}

Obviously, $T$ is always a positive number, and the downlink coverage depends on the SINR threshold $T$. For a certain service type with a sufficiently low threshold, a BS can cover a large area. In contrast, a higher threshold $T$ leads to a smaller coverage area. However, in the ORP scheme, the coverage performance also significantly depends on the number of precoding vectors $N$. In the following theorem, we present the main result of this work, an exact analytical expression for the downlink coverage probability of the ORP scheme. 

\begin{theorem}
\label{theor:Pdl_closeform}
In a system employing the ORP scheme with multiple precoding vectors and a single antenna receiver (ORP-SA scheme), the downlink coverage probability is given by
\begin{align}
\label{P_DL}
\Pdl= 
\begin{cases}
	\frac{N}{\left(T+1\right)^{N-1}} e^{-\frac{TN}{\rho}}, &T \geq 1\\
	\Pdl_1 + \sum_{k=2}^{m-1} \Pdl_k + \Pdl_m, &T < 1,
\end{cases}
\end{align}
where $\Pdl_1$, $\Pdl_k$, and $\Pdl_m$ are
\begin{align*}
	\Pdl_1 	&= \frac{N}{(N-2)!} \left(e^{-\frac{TN}{\rho}} C_1 + C_2\right), \numberthis \label{Pdl_1}\\
	\Pdl_k 
	&= \xi_k \left(e^{-\frac{TN}{\rho}} \frac{T^l}{l!} D_1 - \frac{1}{(k-1)^l l!} D_2\right) \nonumber \\
	&\mathrel{\phantom{=}} + \xi_k \left(\frac{1}{k^l l!} E_1 - \frac{1}{(k-1)^l l!} E_2\right), \numberthis \label{Pdl_k} \\
	\Pdl_m 	&= \xi_m \left(e^{-\frac{TN}{\rho}} \frac{T^l}{l!} F_1 - \frac{1}{(m-1)^l l!}F_2\right). \numberthis \label{Pdl_m}
\end{align*}

Here, the function $\xi_p(\cdot)$, $p = 1, \ldots, N-1$, is defined as
\begin{align}
\label{xi_f}
	\xi_p(\cdot) 
	&= \frac{N}{(N-2)!} \sum_{t=1}^{k} {N-1 \choose t-1} \nonumber \\
	&\mathrel{\phantom{=}} \times (-1)^{t+1} \sum_{i=0}^{N-2} {N-2 \choose i} (1-t)^i i! \sum_{l=0}^{i}(\cdot),
\end{align}
and $C_1$, $C_2$, $D_1$, $D_2$, $E_1$, $E_2$, $F_1$, and $F_2$ are given by \eqref{C_1}--\eqref{F_2} in Appendix \ref{apd:CDEFGH}.
\end{theorem}

\begin{proof}
See Appendix \ref{apd:proof_Pdl}.
\end{proof}

As expected, Theorem \ref{theor:Pdl_closeform} shows that the value of $\Pdl$ varies depending on the SINR threshold $T$ and average SNR $\rho$. Specifically, $\Pdl$ decreases with $T$ but increases with $\rho$. Furthermore, $\Pdl$ depends on $N$, the number of beams. In the ORP scheme, when the number of beams is large, the precoding diversity gain is enhanced. This increases the chances that a precoding vector out of $N$ randomly generated orthogonal ones matches well with the channel of a user to provide high receive signal power. However, at the same time, the effective beam at an MS is affected by additional interference signals introduced by the other beams. Therefore, it is uncertain whether employing a larger number of precoding vectors leads to a better coverage performance. In Remarks \ref{rm:Pdl_approaches_0}--\ref{rm:varying_property_T_smaller_than_1}, the dependencies of $\Pdl$ on $T$ and $N$ in the ORP scheme are stated.
\begin{remark}
	\label{rm:Pdl_approaches_0}
	For any value of $T$, as $N$ increases, the coverage probability approaches zero, i.e., 
	$$
	\Pdl \longrightarrow 0,\text{ as } N \longrightarrow \infty, N \leq N_t, \hspace{0.1cm}\forall T.
	$$
\end{remark}
\begin{proof}
	We observe that
	\begin{align*}
		T\left(\frac{N}{\rho}+b\right) &\longrightarrow \infty, \text{ as } N \longrightarrow \infty, \hspace{0.1cm} \forall T.
	\end{align*}
	Hence, from the expression for the downlink coverage probability in \eqref{I}, we have
	\begin{align*}
		\Pdl 
		&= \int_0^{\infty} \int_{T\left(\frac{N}{\rho}+b\right)}^{\infty} \pdf_{\Am, \Bm}(a,b)dadb \longrightarrow 0, \\
		&\mathrel{\phantom{=}} \hspace{4cm}\text{ as } N \longrightarrow \infty, \hspace{0.1cm} \forall T,
	\end{align*}
	which proves Remark \ref{rm:Pdl_approaches_0}.
\end{proof}
\begin{remark}
\label{rm:varying_property_T_greater_than_1}
When $T\geq1$, the downlink coverage probability is a decreasing function of $N$. Let $N^*$ denote the optimal number of precoding vectors such that the ORP scheme provides the maximum coverage probability. When $T \geq 1$, the maximum coverage probability becomes
$$ 
\Pdl = e^{-\frac{T}{\rho}},
$$
which is achieved for $N^{*}=1$\footnotemark[1]. Furthermore, for multiple precoders, i.e., $N \geq 2$, the higher $N$, the more slowly $\Pdl$ decreases.
\end{remark}

\footnotetext[1]{This does not mean that only a single user is served by the entire system. Multiple users can be simultaneously served with multiple time-frequency resources.}

\begin{proof}
See Appendix \ref{apd:proof_rmk2}.
\end{proof} 

Next, we consider the case of $T < 1$. From \eqref{I}, we see that the downlink coverage probability is determined by $\pdf_{\Am,\Bm}(a,b)$ in \eqref{JoindPdfAmBm} in the area 
\begin{align*}
	\mathcal{R} = \left\{(a,b) \in \mathbb{R}^2_+: T\left(\frac{N}{\rho} + b \right) \leq a \right\}.
\end{align*}
It can be seen from Fig. \ref{fig:R2} in Appendix \ref{apd:proof_Pdl} that when $N$ increases,  sector $\mathcal{R}$ narrows.  However, $\pdf_{\Am,\Bm}(a,b)$ varies depending on $N$. Therefore, in contrast to the case of $T \geq 1$, the decreasing property is not generally secured. If $\Pdl$ is not a decreasing function of $N$, a larger $N^*$ can achieve the maximum $\Pdl$. These properties of $\Pdl$ are stated in the following remark and justified by simulation results in Section \ref{sec:Numerical Results}.

\begin{remark}
	\label{rm:varying_property_T_smaller_than_1}
	When $T < 1$, the conclusion in Remark \ref{rm:varying_property_T_greater_than_1} on the decreasing property of the downlink coverage probability is not valid anymore; thus, the optimal value $N^{*}$ can be larger than one. However, even when $T < 1$, the maximum coverage is achieved for a small number of precoding vectors, i.e. $N^* \ll N_t$. 
\end{remark}

 Remarks \ref{rm:varying_property_T_greater_than_1} and \ref{rm:varying_property_T_smaller_than_1} show that when a sufficiently small number of precoding vectors are employed, the coverage probability becomes higher than when $N$ is large. Furthermore, Remark \ref{rm:Pdl_approaches_0} shows that with a sufficiently large $N$, the coverage probability becomes almost zero. Especially when $T \geq 1$, the downlink coverage probability is a decreasing function of $N$, which results in $N^*=1$. Furthermore, because the coverage probability decreases more rapidly for small $N$, a slight increase of $N$ can substantially lower the coverage probability. These remarks imply that the interference caused by the other beams affects the coverage performance more significantly than the precoding diversity gain. Therefore, for a small number of beams, we achieve the optimal downlink coverage probability, but the number of simultaneously served users is limited. The inverse relationship between the cell coverage and number of active users is also mentioned in \cite{TheVeeravalli}. To overcome this problem, in next sections, MSs equipped with an antenna selection (AS) receiver are considered.
 
 
 Note that based on Theorem \ref{theor:Pdl_closeform}, we can readily derive the cumulative distribution function (CDF) of $X_{max} = \max\limits_{n=1,\ldots,N} SINR_n$ in a baseline system where a single antenna is employed at the receiver. 
 \begin{corollary}
 	\label{corol:cdf_Xmax}
 	The CDF of the random variable $X_{max} = \max\limits_{n=1,\ldots,N} SINR_n$ is given as
 	\begin{align}
 	\label{cdf}
 	\cdf_{X_{max}} (x)= 
 	\begin{cases}
 	1 - \frac{N}{\left(x+1\right)^{N-1}} e^{-\frac{xN}{\rho}}, &x \geq 1\\
 	1 - \left(\Pdlx_1 + \sum_{k=2}^{m-1} \Pdlx_k + \Pdlx_m\right), &x < 1,
 	\end{cases}
 	\end{align}
 	where $\Pdlx_1$, $\Pdlx_k$, and $\Pdlx_m$ are given in \eqref{Pdl_1}, \eqref{Pdl_k}, and \eqref{Pdl_m}, respectively. 
 \end{corollary}
 \begin{proof}
 	By Definition \ref{def:Pdl}, we have
 	\begin{align*}
 	\Pdl 
 	= \prb \left\{ X_{max} > T\right\} = 1 - \cdf_{X_{max}} (T),
 	\end{align*}
 	which leads to
 	\begin{align}
 	\label{cdfxmax}
 	\cdf_{X_{max}} (T) = 1 - \Pdl.
 	\end{align}
 	From \eqref{P_DL} and \eqref{cdfxmax}, we obtain the CDF of $X_{max}$ in \eqref{cdf}.
 \end{proof}
 
 We note that in \cite{OnSharif}, although the ORP scheme is applied, the multiuser diversity gains are exploited, which is different from our scheme. Thus, the maximum SINR among $K$ users is expressed by $X'_{max} = \max\limits_{k=1,\ldots,K} SINR_k$,  and the CDF of $X'_{max}$ is given by \cite{OnSharif}
 \begin{align*}
 \cdf_{X'_{max}}(x) = \left(1 - \frac{e^{-x\frac{N}{\rho}}}{(x+1)^{N-1}}\right)^K.
 \end{align*}
 However, in \cite{OnSharif}, the number of precoding vectors is equal to the number of transmit antennas, i.e., $N = N_t$. This is not practical in massive MIMO systems with the aim of cell coverage extension, because, with a very large number of precoding vectors, the downlink coverage probability tends to zero. This result is presented in Remark \ref{rm:Pdl_approaches_0}.

\subsection{Receivers with AS}
In an AS receiver, multiple receive antennas are utilized to achieve receive spatial diversity gains. The downlink coverage probability of an AS receiver in a system employing the ORP scheme is given in the following theorem.
\begin{theorem}
	\label{theor:Pdl_AS}
	The downlink coverage probability of the ORP scheme with an AS receiver (ORP-AS) is 
	\begin{align}
	\label{Pdlas}
		\Pdl_{AS} = 
		\begin{cases}
			1 - \left(1 - \frac{N}{\left(T+1\right)^{N-1}} e^{-\frac{TN}{\rho}}\right)^{N_r},  \hspace{1.5cm} T \geq 1\\
			1 - \left[1 - \left(\Pdl_1 + \sum_{k=2}^{m-1} \Pdl_k + \Pdl_m\right)\right]^{N_r}, \\
			\hspace{6cm} T < 1,
		\end{cases}
	\end{align}
	where $\Pdl_1$, $\Pdl_k$, and $\Pdl_m$ are given in \eqref{Pdl_1}, \eqref{Pdl_k}, and \eqref{Pdl_m}, respectively. 
\end{theorem}

\begin{proof}
	 Let $X_{max}^{AS}$ denote the maximum SINR in the ORP-AS scheme, i.e.,
	 \begin{align*}
	 	X_{max}^{AS} = \max\limits_{\substack{n=1,\ldots,N \\ r=1,\ldots, N_r}} SINR_{n,r},
	 \end{align*}	 
	  where $SINR_{n,r}$ is the SINR for the $n$th beam at the $r$th antenna of the AS receiver. Because the channel between each pair of transmit and receive antennas are statistically independent, the SINRs at different receive antennas are i.i.d. random variables. Therefore, the CDF of $X_{max}^{AS}$ is given as
	\begin{align}
		\label{cdf_AS}
		\cdf_{\Xm^{AS}} (T) =  \left[\cdf_{X_{max}}(T)\right]^{N_r}.
	\end{align}
	Hence, we obtain
	\begin{align}
		\label{PdlAS_proved}
		\Pdl_{AS} &= 1 - \prb\left\{\Xm^{AS} > T\right\} = 1 - \cdf_{\Xm^{AS}} (T) \nonumber \\
		&= 1- \left[\cdf_{X_{max}}(T)\right]^{N_r}.
	\end{align}
	From \eqref{cdf} and \eqref{PdlAS_proved}, the theorem is proved.
\end{proof}

From Theorem \ref{theor:Pdl_AS}, it is clear that for a fixed $T$ and $\rho$, $\Pdl$ depends on not only on $N$ but also on $N_r$. Larger number of receive antennas mean that more spatial diversity gains can be exploited, which can increase the maximum SINR; hence, higher $\Pdl$ is expected. In the following remark, the coverage performance improvement of the ORP-AS scheme and its dependence on $N$ and $N_r$ are presented.

\begin{remark}
\label{rm:PdlAS_increase_with_Nr}
In the ORP-AS scheme, the downlink coverage probability is an increasing function of $N_r$. In particular, in massive MIMO systems with the ORP-AS scheme that employ a fixed number of precoding vectors, the user is in coverage with a high probability provided that the system is equipped with a sufficiently large number of antennas, i.e., 
\begin{align*}
\Pdl_{AS} \longrightarrow 1 \text{ as } N_r \longrightarrow \infty, N=c.
\end{align*}
where $c$ represents a constant.
\end{remark}
\begin{proof}
With a fixed value of $N$, we observe that 
\begin{align*}
	0 < \cdf_{X_{max}}(T) < 1,
\end{align*}
which leads to
\begin{align*}
	\left[\cdf_{X_{max}}(T)\right]^{N_r} < \cdf_{X_{max}}(T),
\end{align*}
 and $\left[\cdf_{X_{max}}(T)\right]^{N_r}$ is a decreasing function of $N_r$. Therefore, $\Pdl_{AS}$ in \eqref{PdlAS_proved} becomes an increasing function of $N_r$, and $ \Pdl_{AS} \longrightarrow 1, \text{ as } N_r \longrightarrow \infty.$
Remark \ref{rm:PdlAS_increase_with_Nr} is hence proved.
\end{proof}

Remark \ref{rm:PdlAS_increase_with_Nr} reveals that multiple receive antennas enhance the number of transmitted streams as well as the coverage gain. However, according to Remarks  \ref{rm:varying_property_T_greater_than_1} and \ref{rm:varying_property_T_smaller_than_1}, to achieve the maximum downlink coverage probability, the number of precoding vectors should be sufficiently small. Hence, it comes at the price of limiting the number of streams. Fortunately, according to Remark \ref{rm:PdlAS_increase_with_Nr}, when a large number of receive antennas are employed, the BS can transmit multiple streams while the coverage is secured. However, increasing $N$ slows down the increases in $\Pdl_{AS}$, which implies that in the ORP-AS scheme, the cell coverage is still affected by the interference from other beams. In Section \ref{sec:Numerical Results}, we numerically prove the accuracy of this analysis. The coverage can be further improved by employing multiple transmission slots, which is considered in the next section.

\section{ORP with Multiple Precoder Groups Over Multiple Time Slots}
\label{Delay Constraint}
\subsection{Downlink Coverage Probability}
\label{subsec:MP}
The results in the previous sections were obtained by considering a single transmission slot, which consists of a pair of training and transmission phases. In this section, multiple transmission slots are considered to enhance the maximum SINR of each user, which results in a higher cell coverage probability. In this scheme, the BS randomly generates a precoding matrix $\p \in \set^{N_t \times (ND)}$, where $D$ is the number of transmission slots and $\p$ consists of $ND$ orthonormal precoding vectors $\left\{\bp_{n,d}\right\}$, $n=1,\ldots,N$, $d=1,\ldots,D$, with the constraint $ND \leq N_t$. For each transmission slot, one cycle of training and transmission phases occurs.

In the training phase of the first cycle, $N$ training signals are multiplied by the first $N$ precoder group $\left\{\bp_{1,1},\bp_{2,1},\ldots, \bp_{N,1}\right\}$ before being sent to the MSs. Then, each MS computes $N$ SINR values and determines the maximum one. The maximum SINR and the index of the effective beam of each MS are fed back to the BS. In the transmission phase, if an MS has a maximum SINR that is higher than $T$, it is determined to be in coverage and can be selected for transmission. In the second cycle, the next precoder group $\left\{\bp_{1,2},\bp_{2,2},\ldots, \bp_{N,2}\right\}$ is used for precoding. 

We assume that an MS has a delay constraint of $D$ transmission slots for a certain traffic type. In this case, $D$ consecutive cycles of training and transmission phases can be considered to find its effective beam. In this case, the maximum SINR is searched for over $ND$ beams. Therefore, higher $D$ leads to more chances for the maximum SINR to be larger than the threshold, which increases the coverage probability. In the following theorem, we investigate the ORP schemes with single-antenna (SA) and multiple-antenna receivers in conjunction with multiple precoder groups, which are denoted as ORP-MPG and ORP-AS$\&$MPG, respectively.
\begin{theorem}
For transmissions of $D$ multiple precoder groups over multiple time slots, the downlink coverage probabilities of the ORP-MPG and the ORP-AS$\&$MPG schemes are given by
\begin{align*}
	\Pdl_{MPG} &= 
	\begin{cases}
		1 - \left(1 - \frac{N}{\left(T+1\right)^{N-1}} e^{-\frac{TN}{\rho}}\right)^{D}, \hspace{1.1cm} T \geq 1\\
		1 - \left[1 - \left(\Pdl_1 + \sum_{k=2}^{m-1} \Pdl_k + \Pdl_m\right)\right]^{D}, \\
		\hspace{5.5cm} T < 1,
	\end{cases} \numberthis \label{PdlMP}
	\\
	\Pdl_{AS\&MPG} &= 
	\begin{cases}
		1 - \left(1 - \frac{N}{\left(T+1\right)^{N-1}} e^{-\frac{TN}{\rho}}\right)^{N_rD}, \hspace{0.7cm}  T \geq 1\\
		1 - \left[1 - \left(\Pdl_1 + \sum_{k=2}^{m-1} \Pdl_k + \Pdl_m\right)\right]^{N_rD}\\
		\hspace{5.5cm} T < 1,
	\end{cases} \numberthis \label{PdlMPAS}
\end{align*}
respectively, where $\Pdl_1$, $\Pdl_k$, and $\Pdl_m$ are given in \eqref{Pdl_1}, \eqref{Pdl_k}, and \eqref{Pdl_m}, respectively.
\end{theorem}

\begin{proof}
At a receiver, the maximum SINR is selected not only from $N$ beams but also from $D$ transmission slots. In the $d$th slot, the SINR for the $n$th beam, $\bp_{n,d}$, can be expressed as
\begin{align*}
SINR_{n,d} &= \frac{\left|  \h \bp_{n,d}\right|^2 }{\frac{N}{\rho} +  \sum_{i \neq n}^N \left| \h \bp_{i,d}\right| ^2}, \\
&\hspace{3cm} n=1,\ldots, N,\hspace{0.1cm} d=1,\ldots, D.
\end{align*}
In this scheme, the $ND$ precoding vectors in $\p$ are mutually orthogonal, while the coefficients of the channel matrix $\h$ are random variables of $\mathcal{CN}(0,1)$.  Therefore, $\h \bp_{n,d}$, $n=1,\ldots,N$, $d=1,\ldots,D$, are independent. As a result, the $ND$ values of the SINR are independent, which yields
\begin{align*}
\Pdl_{MP} 
&= 1 - \prb \left\{ \max\limits_{\substack{n=1,\ldots,N \\ d=1,\ldots, D}} SINR_{n,d} \leq T \right\} \\
&= 1 - \left[\cdf_{\Xm}(T)\right]^{D} \numberthis \label{Pdl_dc}.
\end{align*}
From \eqref{cdf}, \eqref{cdfxmax}, and \eqref{Pdl_dc}, we obtain $\Pdl_{MP}$ in \eqref{PdlMP}. By performing the same analysis as for the case of multiple receive antennas with the AS receiver, we complete the proof of this theorem.
\end{proof} 

We note that a similar analysis can be performed for broadcast channels, which do not require the training phases. In broadcast channels, when the SINR of a beam exceeds the threshold, the signal received over the corresponding beam can be successfully decoded, which means that the MS is in coverage. If a broadcast channel has a delay constraint of $D$, the ORP scheme for this channel provides the same coverage probabilities as \eqref{PdlMP} and \eqref{PdlMPAS}.

\subsection{Comparison of ORP and STC}
\label{sec:STC_and_ORP}
In this section, we compare a massive MIMO system using the ORP scheme with one using the STC technique. STC improves the reliability of the link by achieving a high diversity order. Furthermore, the STC scheme generally does not exploit the CSI at the transmitter, and hence, no CSI feedback is required. Therefore, the STC is widely considered to be a suitable transmission scheme for transmission to cell-edge users as well as the transmission of broadcast/multicast signals. As an example, in long-term evolution mobile networks, the STC is employed for the physical broadcast channel, which should be received by all the users in a cell \cite{etsi36211}.

For the sake of simplicity, we assume that only one receive antenna is employed, i.e., $N_r=1$, and we consider a real space-time block code $\textbf{\textit{x}} \in \setR^{N_t \times 1}$ as the transmit signal. In STC, the effective received signals at a user can be written as
\begin{align}
\label{STC}
\textit{\textbf{y}} = \sqrt{\frac{E_x}{\sigma^2 N_t}} \h \textit{\textbf{x}} + \textit{\textbf{z}},
\end{align} 
where $E_x$ and $\sigma^2$ indicate the average energy of each transmit signal and the noise variance, respectively. Furthermore, $\textit{\textbf{y}} \in \set^{N_t \times 1}$, $\h \in \set^{N_t \times N_t},$ and $ \textit{\textbf{z}} \in \set^{N_t \times 1} $ denote the receive signal vector, channel matrix, and Gaussian noise vector, respectively. In \eqref{STC}, \textit{\textbf{y}} is obtained by properly rearranging the original received signals, and the columns of $\h$ are orthogonal to each other \cite{cho2010mimo}. Because of the orthogonality of $\h$, the signals can be optimally detected by a linear receiver, which yields
\begin{align*}
\tilde{\textbf{\textit{y}}} 
&= \h^T \textbf{\textit{y}}
= \sqrt{\frac{E_x}{\sigma^2 N_t}} \h^T \h \textit{\textbf{x}} + \h^T \textbf{\textit{z}} 
=  \sqrt{\frac{E_x}{\sigma^2 N_t}} \sum_{i=1}^{N_t} \left| h_i\right|^2 \textbf{\textit{x}} + \tilde{\textbf{\textit{z}}}.
\end{align*}

Note that with the STC scheme in massive MIMO systems, a very large number of transmit antennas hardens the time-varying channel \cite{OpportunisticViswanath}. By the law of large numbers, we have
\begin{align*}
\lim_{N_t \longrightarrow \infty} \frac{1}{N_t} \sum_{i=1}^{N_t} \left| h_i\right|^2 \longrightarrow 1.
\end{align*}
As a result, a fading channel in a massive MIMO system turns into a less varying channel \cite{al2010opportunistic}. With the STC scheme, because the channel is averaged over $N_t$ values, the SNR at the receivers varies less in the time domain. Hence, for the downlink coverage probability, it gains little diversity over $D$ transmission slots. When only one transmission slot is employed, it is more likely that the STC technique will provide a better coverage compared to the ORP scheme because it has no inter-stream interference. However, when signals are transmitted over multiple slots, in the ORP scheme, a higher diversity gain can be achieved by multiple precoder groups. As explained in Section \ref{subsec:MP}, if the signals are transmitted over $D$ slots, the coverage probability is determined by the maximum SINR over $ND$ values. Therefore, when $D$ is sufficiently large, it is expected that the maximum SINR of the ORP scheme will be higher than the average SNR of the STC scheme. As a result, the ORP scheme can outperform the STC scheme when multiple transmission slots are employed, which is verified through numerical results in Section \ref{sec:Numerical Results}.

We also compare the two schemes in terms of the overhead for pilot signals. In the STC scheme, the channel coefficient of each transmit antenna should be estimated for symbol detection, and hence the overhead of pilot signals increases proportionally with $N_t$. In particular, in the massive MIMO system, the overhead for pilot signals becomes significant, which can substantially limit the spectral efficiency. In contrast, the number of required pilot signals increase proportionally with the number of beams, which should be set to a small number to maximize the coverage, as explained in Section \ref{sec:Downlink Coverage Probability with ORP}. Therefore, the ORP scheme requires a significantly lower number of pilot symbols for the signal transmission to cell-edge users while being able to enhance their link quality.
\section{Numerical Results}
\label{sec:Numerical Results}
Computer simulations were performed to evaluate the performance of the proposed ORP scheme. An orthonormal precoding matrix is created by computing an orthonormal basis for the column space of a randomly generated matrix. Furthermore, the coefficients of the channel matrix are randomly generated as $\mathcal{CN}(0, 1)$ random variables but fixed over $D$ time slots. The  values of $N_t$, $N_r$, $N$, $D$, $\rho$, and $T$ are differently assumed in each simulation.

\begin{figure}[t]
	\centering
	\includegraphics[scale=0.65]{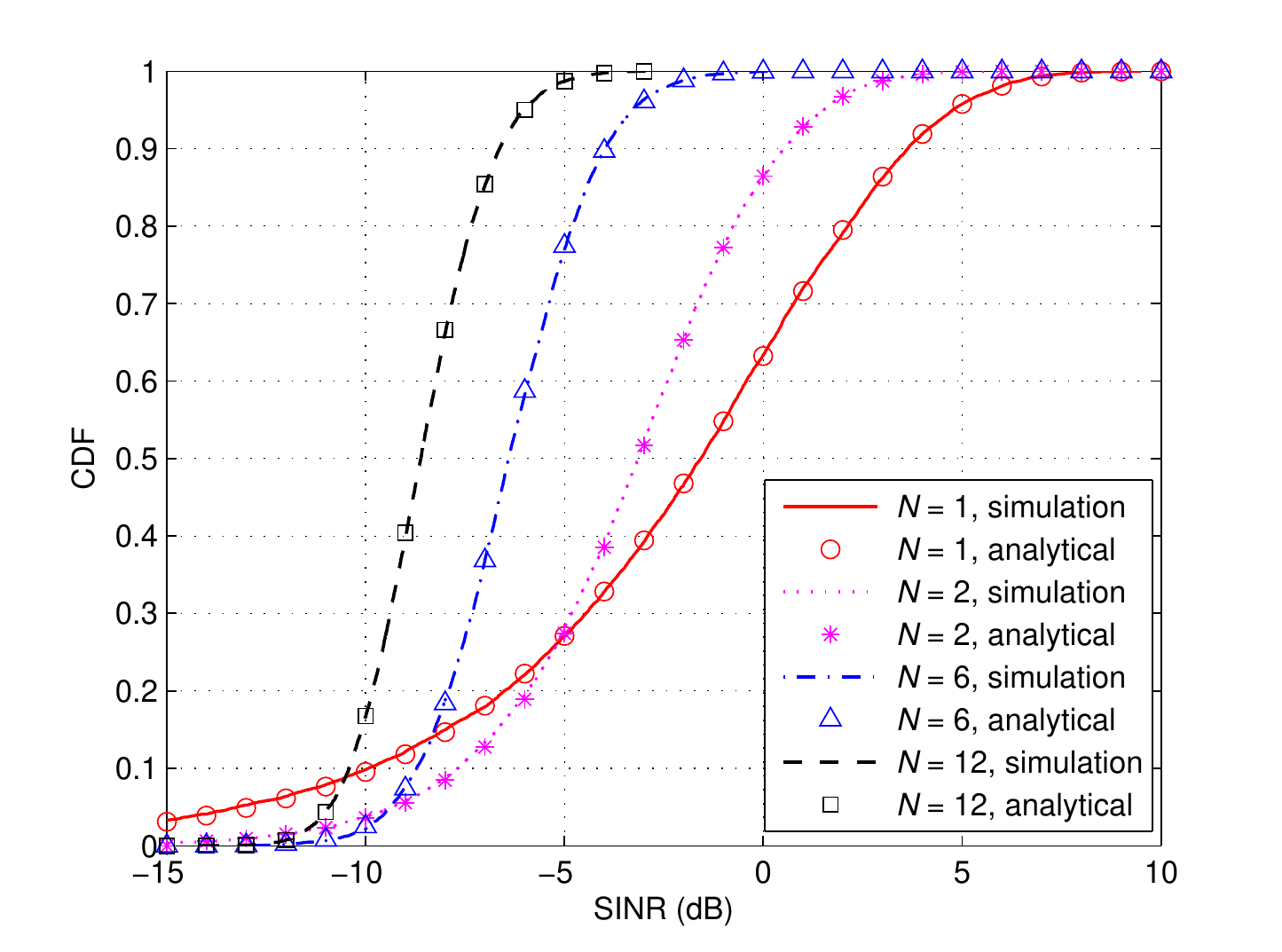}
	\caption{Comparison between analytical and simulation CDFs of the maximum SINR for $\rho = 0$ dB, $N_t = 32$, $N_r = 1$, and $N \in \{1,2,6,12\}$.}
	\label{fig:CDF_SINR_max}
\end{figure}

\begin{figure}[t]
 	\centering
 	\includegraphics[scale=0.65]{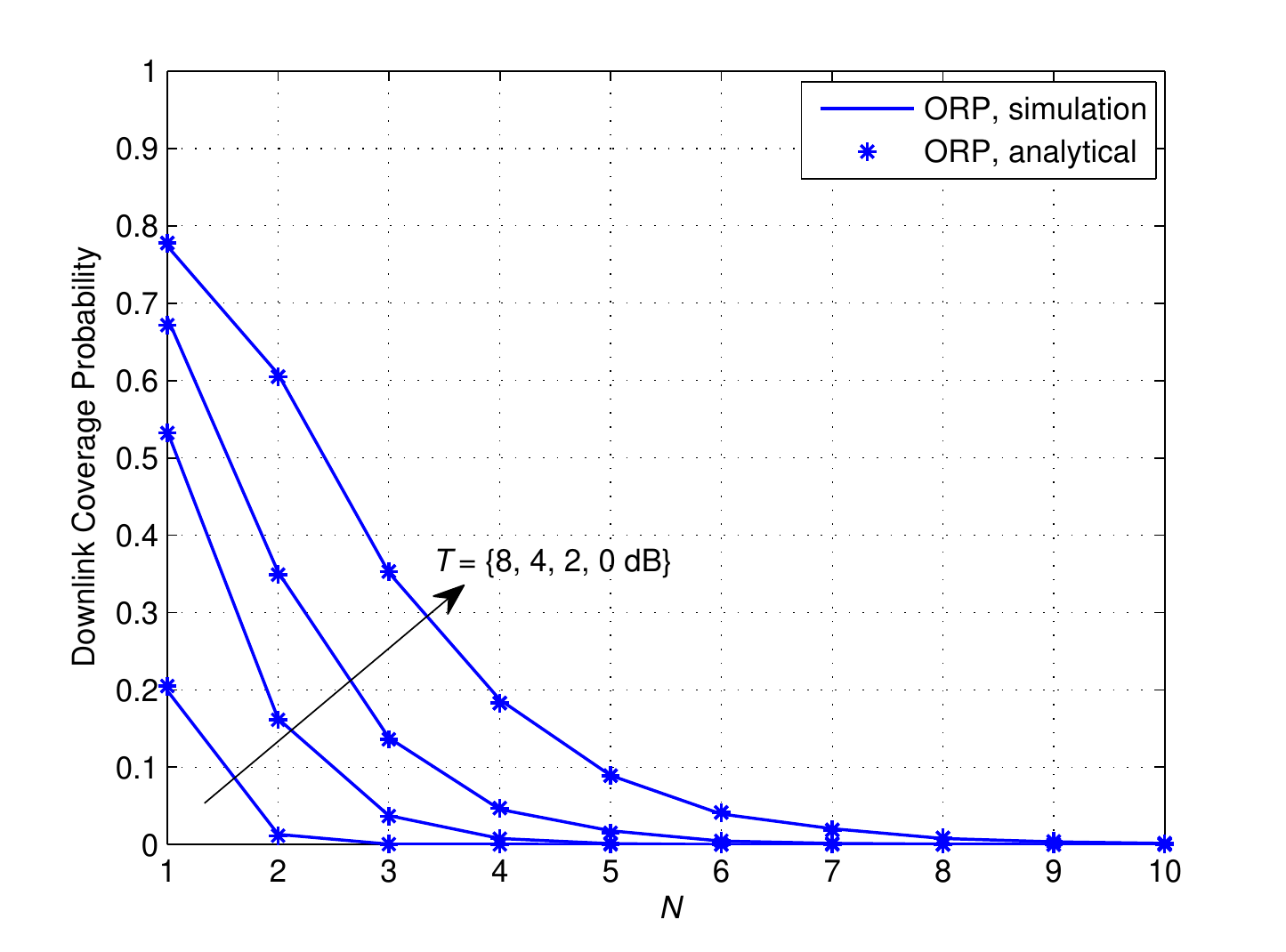}
 	\caption{Downlink coverage probability versus $N$ when $T\geq1$ for $N_t = 32$, $N_r=1$, $\rho = 6$ dB, and $T \in \{0,2,4,8\}$ dB.}
 	\label{fig:PDLwithTgreaterThan1}
 \end{figure}

\begin{figure}[t]
 	\centering
 	\includegraphics[scale=0.65]{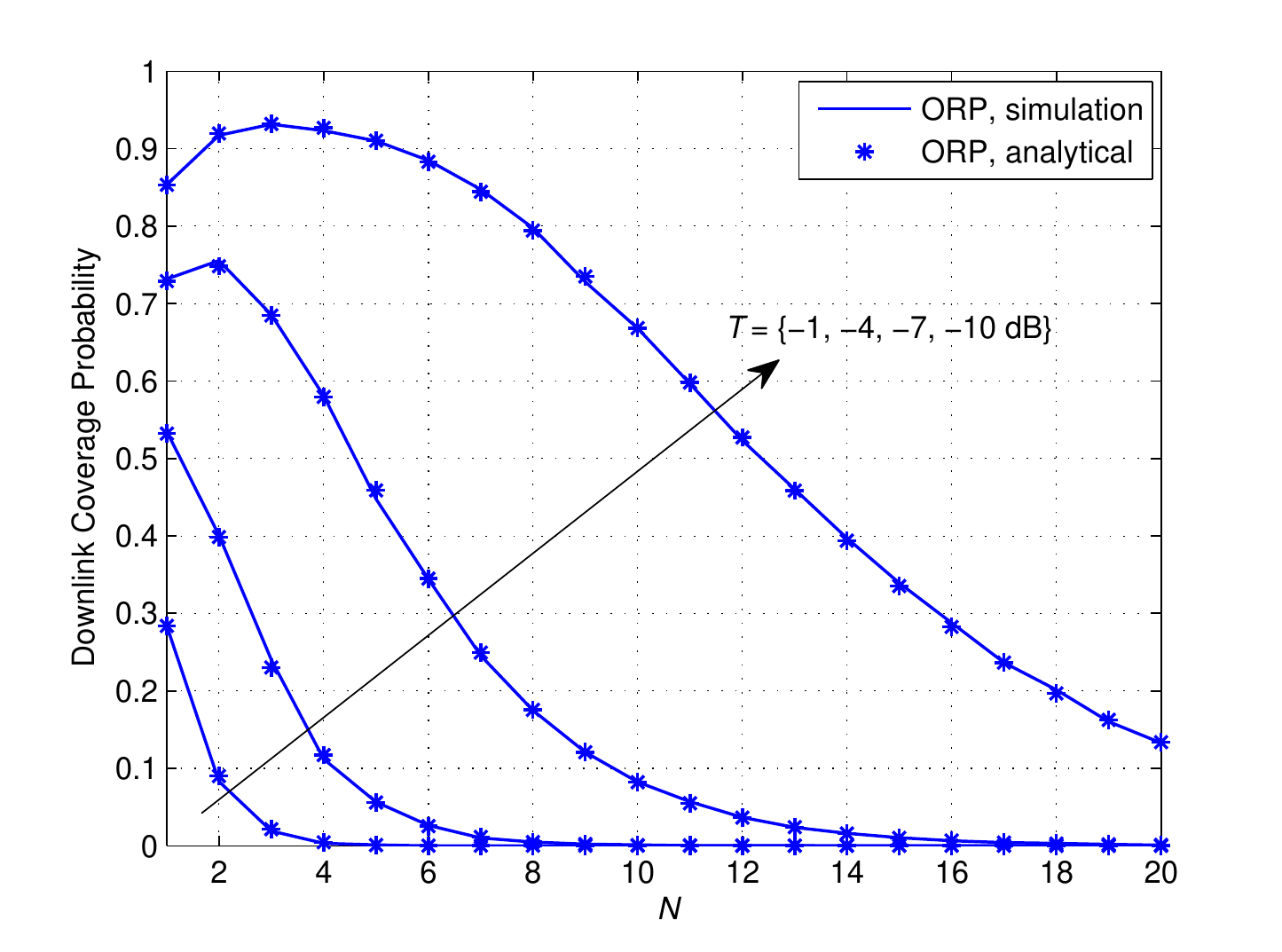}
 	\caption{Downlink coverage probability versus $N$ when $T<1$ for $N_t=32$, $N_r=1$, $\rho = -2$ dB, and $T \in \{-1,-4,-7,-10\}$ dB.}
 	\label{fig:Pdl_vs_T_smaller_than_1}
 \end{figure}

First, we validate the accuracy of the analytical CDF of the maximum SINR in the ORP scheme, which is given in Corollary \ref{corol:cdf_Xmax}, by comparing it to the simulation results. In Fig. \ref{fig:CDF_SINR_max}, the CDFs of the maximum SINR of the ORP scheme for $N_t = 32$, $N_r = 1$,  $\rho = 0$ dB, and $N \in \{1,2,6,12\}$ are depicted. It is clear that the analytical results in \eqref{cdf} match well with the simulation results. We observe that in the low SINR region, the ORP scheme with small $N$ has larger CDF values than the one with larger $N$. However, in the high SINR region, larger $N$ leads to a larger CDF value. Therefore, if the SINR threshold $T$ is low, it is better to use multiple precoding vectors to maximize the cell coverage. In contrast, as $T$ is sufficiently high, only a single precoding vector should be used to maximize the cell coverage, as proved in Remark \ref{rm:varying_property_T_greater_than_1}. This property of the CDF helps explain the results in Figs. \ref{fig:PDLwithTgreaterThan1} and \ref{fig:Pdl_vs_T_smaller_than_1}.

Figs. \ref{fig:PDLwithTgreaterThan1} and \ref{fig:Pdl_vs_T_smaller_than_1} present the results of the downlink coverage probability of the ORP-SA scheme ($N_r=1$) to validate the accuracy of the analytical expression of $\Pdl$ in Theorem \ref{theor:Pdl_closeform}. In Fig. \ref{fig:PDLwithTgreaterThan1}, the case $T \geq 1$ is considered, while Fig. \ref{fig:Pdl_vs_T_smaller_than_1} depicts $\Pdl$ for $T < 1$. In each figure, $\rho$ is fixed, while various values of $T$ are assumed. In Figs. \ref{fig:PDLwithTgreaterThan1} and \ref{fig:Pdl_vs_T_smaller_than_1}, it is clear that the results from the formula in Theorem \ref{theor:Pdl_closeform} agree with those from the simulations. It can also be observed that as $T$ decreases, the coverage performance is significantly improved. Furthermore, as $N$ increases, as stated in Remark \ref{rm:Pdl_approaches_0}, $\Pdl$ approaches zero and becomes substantially smaller than it is for small $N$. It is clear that the optimal number of precoding vectors $N^*$ depends on $T$ and $\rho$.

Specifically, in Fig. \ref{fig:PDLwithTgreaterThan1}, where the higher-than-one SINR threshold, $T \in \{0,2,4,8\}$ dB, $\rho=6$ dB, and $N_t=32$ are considered, it can be observed that the downlink coverage probability is a strictly decreasing function of $N$; thus, $\Pdl$ is always maximum at $N^*=1$, and rapidly decreases to zero as $N$ grows. This result implies that when $T \geq 1$, the coverage performance is seriously affected by the interference from ineffective beams rather than benefiting from the diversity gains. Another observation from Fig. \ref{fig:PDLwithTgreaterThan1} is that for $N \geq 2$, the higher $N$ is, the more slowly $\Pdl$ decreases, as discussed in Remark \ref{rm:varying_property_T_greater_than_1}.

 Fig. \ref{fig:Pdl_vs_T_smaller_than_1} shows the downlink coverage probability for $T<1$. In this simulation, we assume $N_t=32$, $\rho = -2$ dB, and $T \in \{-1,-4,-7,-10\}$ dB. Compared to the results in Fig. \ref{fig:PDLwithTgreaterThan1}, it can be observed that for $T<1$, $\Pdl$ is not generally a decreasing function of $N$. It is interesting to note that when $T$ becomes substantially smaller, the optimal point $N^*$ tends to be a larger value. Therefore, the coverage performance in the massive MIMO downlink with substantially small $T$ can be improved by using multiple precoding vectors. However, in the assumed environments, $N^*$ is not larger than three. Furthermore, after achieving its peak at a relatively small $N$, $\Pdl$ approaches zero, which further proves that the use of a large number of beams is not desirable for optimizing cell coverage.
 
\begin{figure}[t]
	\centering
	\includegraphics[scale=0.65]{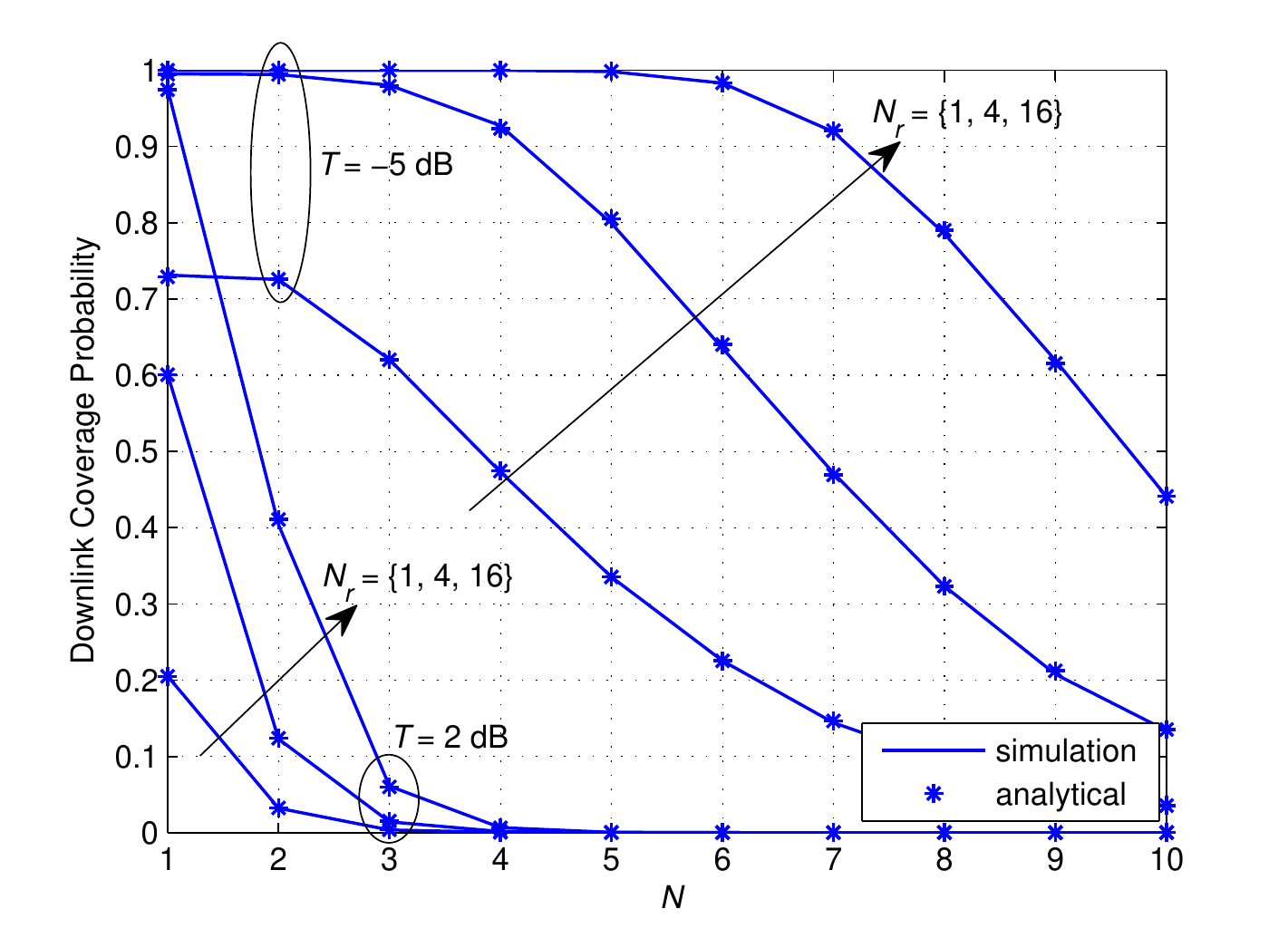}
	\caption{Comparison between ORP-SA and ORP-AS for $N_t=32$, $N_r \in \{1, 4, 16\}$, $\rho = 0$ dB, and $T \in \{-5, 2\}$ dB.}
	\label{fig:AS_vs_Nr}
\end{figure}

\begin{figure}[t]
	\centering
	\includegraphics[scale=0.65]{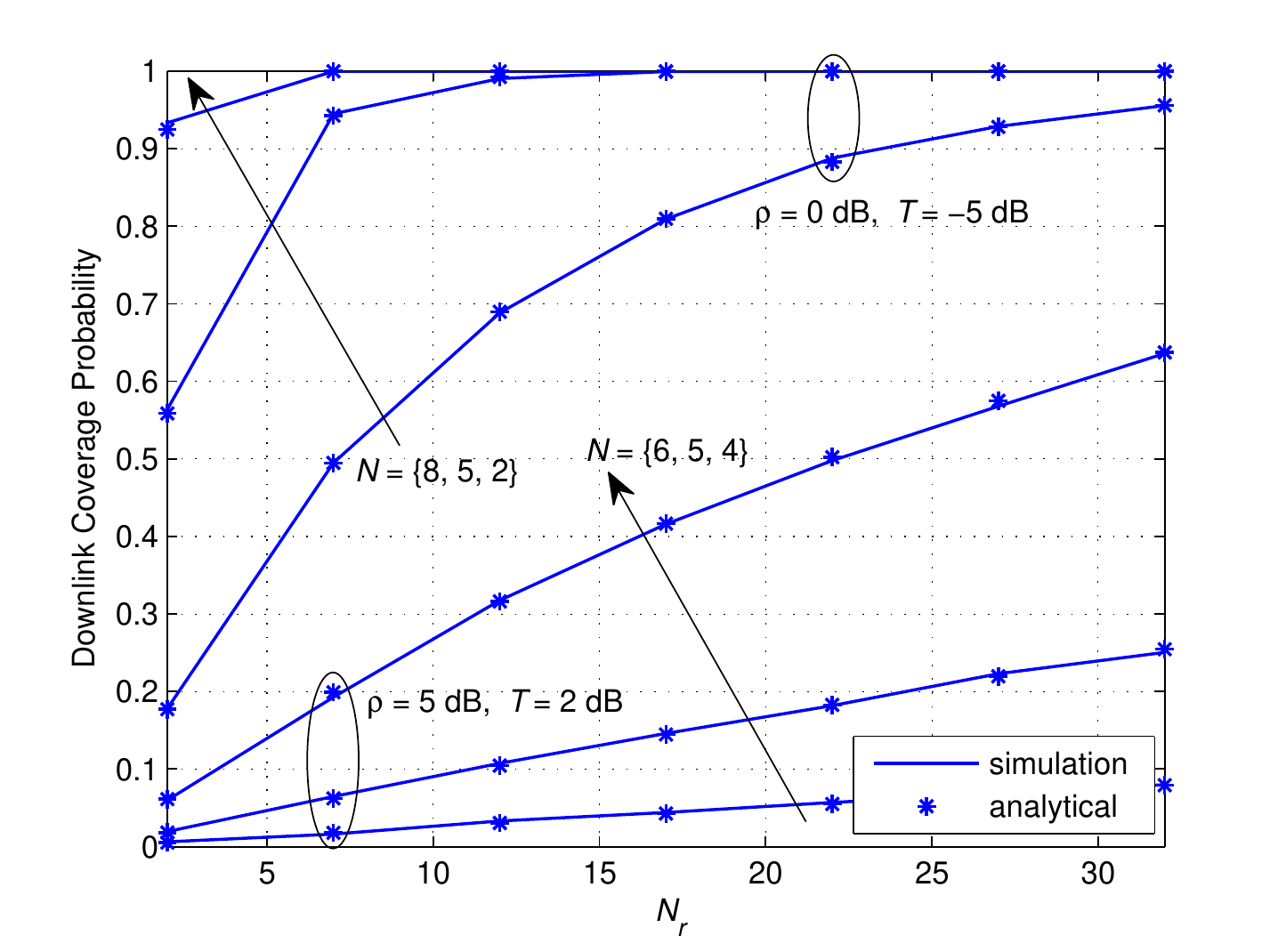}
	\caption{Downlink coverage probability versus $N_r$ in the ORP-AS scheme for $N_t=32$ and $\{\rho, T\} = \left\{\{0, -5\},\{5, 2\}\right\}$ dB.}
	\label{fig:AS_NrOverN}
\end{figure}

 Figs. \ref{fig:AS_vs_Nr} and \ref{fig:AS_NrOverN} present the coverage performance of the ORP-AS scheme to numerically verify Theorem \ref{theor:Pdl_AS} and Remark \ref{rm:PdlAS_increase_with_Nr}. In these figures, for $N_t=32$, the simulation and analytical results of the downlink coverage probability are depicted to show that they match well for all cases of $T$, $N$, and $N_r$. In Fig. \ref{fig:AS_vs_Nr}, we compare $\Pdl$ of the ORP-SA $(N_r=1)$ and ORP-AS schemes $(N_r \in \{4, 16\})$ for $T\in\{-5,2\}$ dB and $\rho=0$ dB. For both values of $T$, it is clear that the ORP-AS scheme provides a significantly better coverage performance than the ORP-SA scheme. For example, for $T=-5$ dB, in the ORP-AS scheme with $N_r=16$ and $N \in [1;6]$, $\Pdl_{AS} \approx 1$ is achieved; however, for the ORP-SA scheme, the maximum coverage probability is only $0.73$ at $N=1$. 

Fig. \ref{fig:AS_NrOverN} shows the results of $\Pdl_{AS}$ versus $N_r$ for various values of $T$, $\rho$, and $N$, when $N_r$ increases from 2 to 32 and $N_t = 32$. Specifically, we consider two cases: $T < 0$, $\{T, \rho\} = \{-2, 0\}$ dB and $T \geq 0$, $\{T, \rho\} = \{2, 5\}$ dB. It can be observed that in both cases, for a fixed $N$, $\Pdl_{AS}$ is an increasing function of $N_r$, and we get $\Pdl_{AS} \longrightarrow 1$ as $N_r \longrightarrow \infty$. For $T=-5$ dB and $\rho = 0$ dB, if the MS is equipped with 12 antennas, five streams can be simultaneously transmitted with $\Pdl_{AS} \approx 1$. This result implies that in the ORP-AS scheme, multiple data streams can be transmitted while preserving high coverage probability. Furthermore, it can be seen from Fig. \ref{fig:AS_NrOverN} that for $T \geq 1$, the rate of increase of $\Pdl_{AS}$ decreases as $N$ increases. These properties agree with the conclusions in Remark \ref{rm:PdlAS_increase_with_Nr}.

\begin{figure}[t]
	\centering
	\includegraphics[scale=0.65]{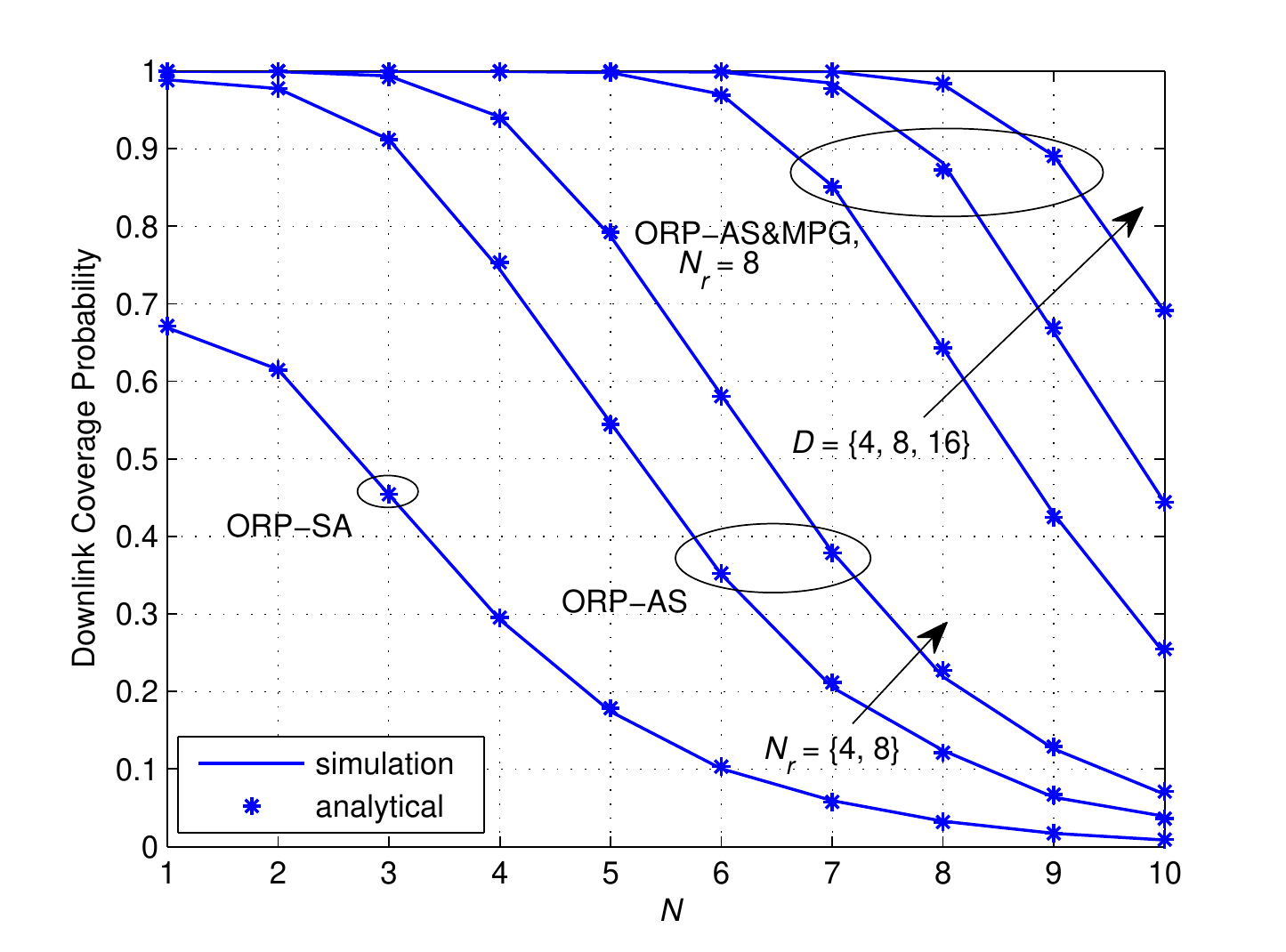}
	\caption{Comparison of the ORP-SA, ORP-AS, and ORP-AS$\&$MPG  schemes for $N_t = 200$, $\rho = 0$ dB, and $T = -4$ dB.}
	\label{fig:Compare_AS_DC_SingleRx}
\end{figure}

We now compare the coverage probabilities of the ORP-SA, ORP-AS, and ORP-AS$\&$MPG schemes. In Fig. \ref{fig:Compare_AS_DC_SingleRx}, both analytical and simulation results are presented for these schemes. In this simulation, $\rho = 0$ dB, $T = -4$ dB, $N_t = 200$, $N_r \in \{4, 8\}$, and $D \in \{4, 8, 16\}$ are assumed. Fig. \ref{fig:Compare_AS_DC_SingleRx} shows that the analytical results agree with the simulation results. It can be seen that in the ORP-AS$\&$MPG scheme, better cell coverage is achieved with higher $D$. It is also clear that the ORP-AS$\&$MPG scheme achieves substantially higher coverage probabilities compared to the ORP-SA and ORP-AS schemes. In other words, the combination of the ORP scheme with an AS receiver and multiple precoder groups significantly improves the coverage performance.

\begin{figure}[t]
	\centering
	\includegraphics[scale=0.65]{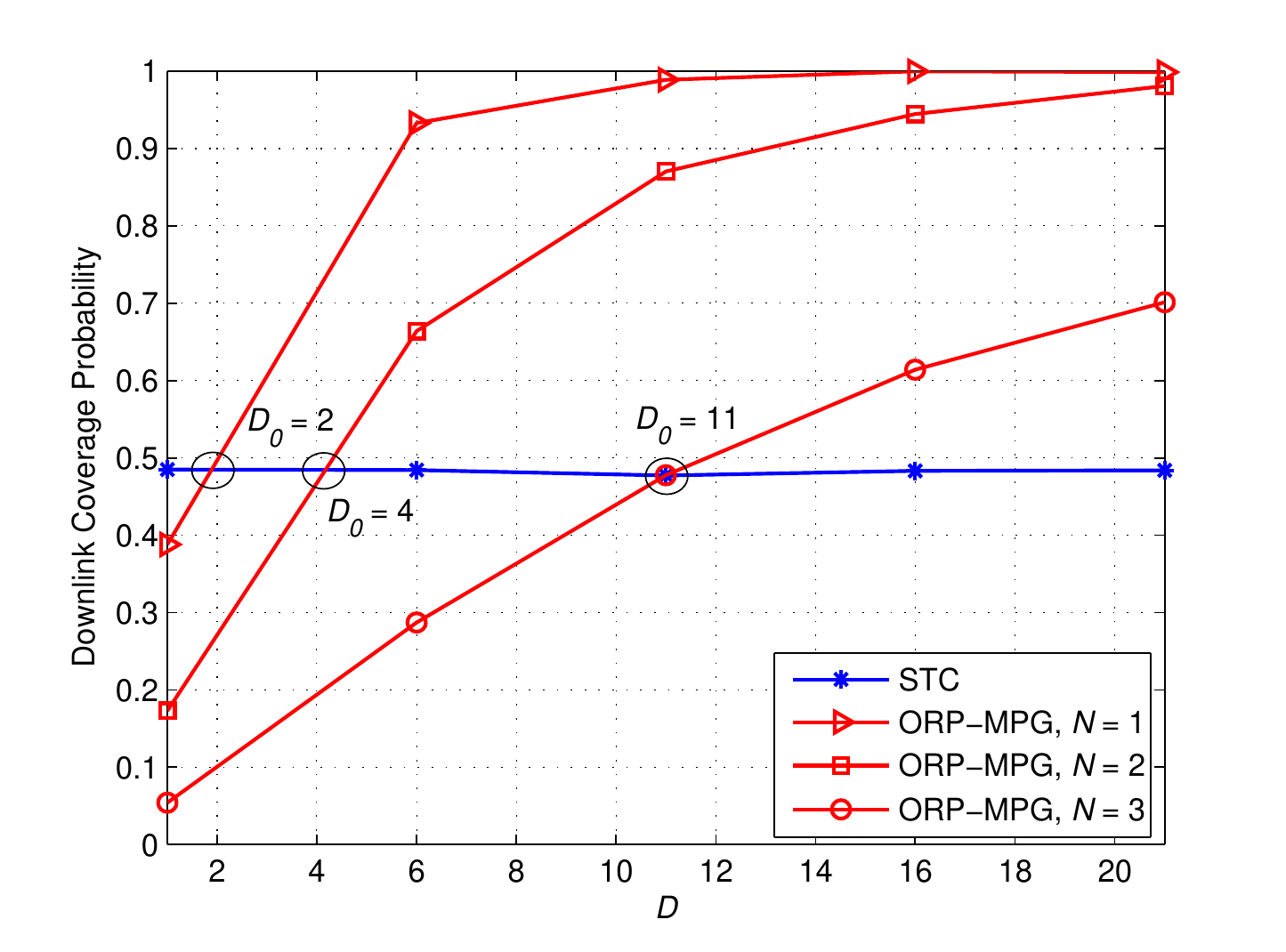}
	\caption{Comparison of the STC and ORP schemes for $N_t = 64$, $N_r = 1$, $N \in \{1, 2, 3\}$, and $T = \rho = -2$ dB.}
	\label{fig:Compare_STC_ORP}
\end{figure}

Finally, we compare the downlink coverage probabilities of the ORP-MPG and STC schemes when  $T = \rho = -2$ dB, where $N_t = 64$, $N_r = 1$, and $N \in \{1, 2, 3\}$. In Fig. \ref{fig:Compare_STC_ORP}, it is shown that as $D$ increases, the coverage probability of the ORP scheme increases. In contrast, the STC scheme has almost constant coverage probability because the channel is assumed to be fixed over $D$ time slots. We define $D_0$ as the reference point where both schemes achieve the closest coverage probability. For example, in the ORP scheme with $N = 1$, $D_0$ is equal to two, which means that with more than two transmission slots, the ORP scheme obtains a higher coverage probability than the STC scheme. Furthermore, we obtain $\Pdl \approx 1$ if more than 11 transmission slots are employed while the STC scheme achieves a coverage probability of 0.48 for every value of $D$. This result shows  that the ORP scheme employing multiple precoder groups over $D$ time slots is capable of providing significant performance gains in terms of coverage over the STC scheme.
\section{Conclusion}
\label{Conclusion}
In this paper, the cell coverage extension problem in massive MIMO systems was considered. As one eligible solution for this problem, we proposed the use of the ORP scheme, where the transmit signals are precoded by the orthonormal precoding vectors. We first analyzed the coverage performance in the downlink when an SA receiver is employed. The analytical closed-form expression of the coverage probability was derived, which shows that the maximum coverage probability not only depends on the SINR threshold $T$, but also significantly depends on the number of precoding vectors $N$. It was also shown that to reduce the deleterious effects of interference from the ineffective beams and to achieve optimal coverage performance, the use of a small number of precoding vectors is desirable. To further extend the coverage, we investigated the ORP-AS, ORP-MPG, and combined ORP-AS$\&$MPG schemes, which can significantly improve the coverage performance. The analytical results were confirmed through numerical results, which proved the accuracy of our derived expressions.

Finally, we compared the proposed ORP scheme to the STC scheme over multiple time slots under a delay constraint. It was numerically shown that the ORP scheme with multiple precoder groups is capable of providing a higher coverage probability than the STC scheme. The analytical and numerical results prove that the ORP scheme can efficiently extend the coverage of unicast channels with a small number of feedback signals. Furthermore, it can also be employed for multicast/broadcast channels, where CSI-based precoding is typically infeasible. Note that, even though the coverage analysis was performed for massive MIMO systems by considering a large number of beams and antennas, Theorems \ref{theor:Pdl_closeform}--\ref{theor:Pdl_AS} and Remarks \ref{rm:varying_property_T_greater_than_1}--\ref{rm:varying_property_T_smaller_than_1} can also be applied to normal MIMO systems where a moderate number of antennas are employed. In future work, the analysis could consider coverage extension with different multiple-antenna receivers such as minimum-mean-square-error, maximum ratio combining, and zero-forcing receivers, which can potentially further increase the cell coverage. {The analysis could also be extended to investigate the coverage extension for the practical multi-cell environment with inter-cell interference.}

\appendices
\section{Expressions for $C_1$, $C_2$, $D_1$, $D_2$, $E_1$, $E_2$, $F_1$, and $F_2$}
\label{apd:CDEFGH}
\begin{align*}
C_1 &= \frac{(N-2)!}{(T+1)^{N-1}} \left(1-e^{-(T+1)b_1} \sum_{l=0}^{N-2} \frac{(T+1)^l b_1^l}{l!}\right), \numberthis \label{C_1}\\ 
C_2 &= \frac{(N-2)!}{2^{N-1}} e^{-2b_1} \sum_{l=0}^{N-2} \frac{2^l b_1^l}{l!},\numberthis \label{C_2}\\
D_1 &= \sum_{v=0}^l {l \choose v} \left(\frac{N}{\rho}\right)^{l-v} \frac{(N-i+v-2)!}{\left(T+1\right)^{N-i+v-1}} \times \\
&\sum_{u=0}^{N-i+v-2}  \frac{\left(T+1\right)^u}{u!}  \left(e^{-(T+1)b_{k-1}}  b_{k-1}^u - e^{-(T+1)b_{k}} b_{k}^u \right),\numberthis \label{D_1} \\
D_2 &= \frac{(N-i+l-2)!}{\left(\frac{k}{k-1}\right)^{N-i+l-1}} \left[e^{-\frac{k}{k-1} b_{k-1}} \sum_{u=0}^{N-i+l-2} \left(\frac{k}{k-1}\right)^u \right.\\
&\mathrel{\phantom{=}} \left. \times \frac{b_{k-1}^u}{u!} - e^{-\frac{k}{k-1} b_{k}} \sum_{u=0}^{N-i+l-2} \left(\frac{k}{k-1}\right)^u \frac{b_{k}^u}{u!} \right],\numberthis \label{D_2}\\
E_1 &= \frac{(N-i+l-2)!}{\left(\frac{k+1}{k}\right)^{N-i+l-1}} e^{-\frac{k+1}{k} b_{k}} \sum_{u=0}^{N-i+l-2} \left(\frac{k+1}{k}\right)^u \frac{b_{k}^u}{u!} ,\numberthis \label{E_1}\\
E_2 &= \frac{(N-i+l-2)!}{\left(\frac{k}{k-1}\right)^{N-i+l-1}} e^{-\frac{k}{k-1} b_{k}} \sum_{u=0}^{N-i+l-2} \left(\frac{k}{k-1}\right)^u \frac{b_{k}^u}{u!}, \numberthis \label{E_2}\\
F_1 &= \sum_{v=0}^l {l \choose v} \left(\frac{N}{\rho}\right)^{l-v} \frac{(N-i+v-2)!}{\left(T+1\right)^{N-i+v-1}}   e^{-(T+1)b_{m-1}}\\
&\mathrel{\phantom{=}} \hspace{2.7cm}\times \sum_{u=0}^{N-i+v-2}  \left(T+1\right)^u   \frac{b_{m-1}^u}{u!} , \numberthis \label{F_1}\\
F_2 &= \frac{(N-i+l-2)!}{\left(\frac{m}{m-1}\right)^{N-i+l-1}} e^{-\frac{m}{m-1} b_{m-1}} \\ 
&\mathrel{\phantom{=}} \hspace{2.5cm} \times \sum_{u=0}^{N-i+l-2} \left(\frac{m}{m-1}\right)^u \frac{b_{m-1}^u}{u!}, \numberthis \label{F_2}
\end{align*}

where $b_k$, $k = 1,\ldots,N-1$, is given as
\begin{align}
\label{bn_point}
b_k = \frac{TN}{\rho (1/k-T)}.
\end{align}


\section{Proof of Theorem \ref{theor:Pdl_closeform}}
\label{apd:proof_Pdl}
	Let $A_n=\left|  \hv^T \bp_{n}\right|^2$, $B_n = \sum_{i \neq n}^N \left| \hv^T \bp_{i}\right| ^2$, and $X_n = SINR_{n}$. We rewrite \eqref{SINR_eqn} as
	\begin{align}
	\label{X_rv}
	X_n = \frac{A_n}{N/\rho + B_n}.
	\end{align}
	We denote $X_{max} = \max\limits_{n=1,\ldots,N} SINR_n$, $\Am = \max\limits_{n=1,\ldots,N} A_n$, and $\Bm = \min\limits_{n=1,\ldots,N}\sum_{i \neq n}^N B_n$. We can write
	\begin{align}
	\label{Xm}
	\Xm = \frac{\Am}{N/\rho + \Bm},
	\end{align}
	where we have $\Am \leq \frac{\Bm}{N-1}$. From Definition \ref{def:Pdl} and \eqref{Xm}, the downlink coverage probability can be expressed as
	\begin{align*}
	\Pdl 
	&= \prb \left\{\Xm > T \right\}\\
	&= \prb\left\{\Am > T\left(\frac{N}{\rho} + \Bm \right)\right\}\\
	&= \int_0^\infty \int_{T\left(\frac{N}{\rho}+b\right)}^{\infty} \pdf_{\Am,\Bm}(a,b)dadb. \numberthis \label{I}
	\end{align*}
	
First, we derive the joint distribution of $\Am$ and $\Bm$, i.e., $f_{\Am, \Bm}(a,b)$. 
	Because $\p$ is composed of orthonormal vectors and the coefficients of $\hv^T$ are random variables of  $\mathcal{CN}(0,1)$, $\hv^T \bp_{n}$ has the same distribution. Therefore, $A_n$ becomes a central chi-square random variable of two degrees of freedom with mean $\epsilon^2 = 1$ , i.e., $A_n \sim \chi^2\left(1\right)$. The probability density function (PDF) and the CDF of $A_n$ are given by
	\begin{align*}
	\pdf_{A_n}(a) &= e^{-a},\\
	\cdf_{A_n}(a) &= 1 - e^{-a}, \numberthis \label{CDF_A}
	\end{align*}
	respectively.
	Because $A_i=\left|  \hv^T \bp_{i}\right|^2$ and $A_j=\left|  \hv^T \bp_{j}\right|^2$ are independent for all $i$ and $j$, we have
\begin{align*}
\cdf_{\Am}(a) 
&= \prb\left\{\Am \leq a\right\} = [\prb\left\{A_n \leq a\right\}]^N  = [\cdf_{A_n}(a)]^N \\
&= (1-e^{-a})^N,
\end{align*}
which leads to the PDF of $\Am$:
\begin{align*}
\pdf_{A_{max}}(a) 
= \frac{d}{da} \cdf_{\Am}(a) = N(1-e^{-a})^{N-1} e^{-a} \numberthis \label{fAm}.
\end{align*}

Let $S = \sum_{n=1}^{N}\left|  \hv^T \bp_{n}\right|^2 = \sum_{n=1}^{N} A_n = \Am + \Bm$.  We observe that
\begin{align*}
\cdf_{\Bm|A_{max}}(b|a) 
&= \prb\left\{\Bm \leq b|A_{max} = a\right\} \\
&= \prb\left\{\Am+\Bm \leq a+b |\Am=a\right\}\\
&= \prb \left\{ S \leq a+b | \Am=a \right\} \\
&= \cdf_{S|\Am}(s|a),  \numberthis \label{CDF_SA}
\end{align*}
where $s=a+b$. The PDF of $S$ conditioned on $\Am$ is expressed as \cite{qeadan2012joint}
\begin{align}
\label{SgA}
\pdf_{S|\Am} (s|a) = \frac{s^{N-2} e^{-(s-a)}}{N! (1-e^{-a})^{N-1}} h_N\left(\frac{a}{s}\right),
\end{align}
where
\begin{align}
	\label{hn}
h_N\left(\frac{a}{s}\right) = 
\begin{cases}
N(N-1) \sum_{t=1}^k {N-1 \choose t-1} (-1)^{t+1} \left( 1 - t\frac{a}{s} \right)^{N-2}, \\
\hspace{2cm}\frac{1}{k+1} \leq \frac{a}{s} \leq \frac{1}{k}, \hspace{0.1cm} k=1,\ldots, N-1 \\
0, \hspace{1.7cm} \text{otherwise}.
\end{cases}
\end{align}
\begin{figure}[H]
	\centering
	\includegraphics[scale=1.0]{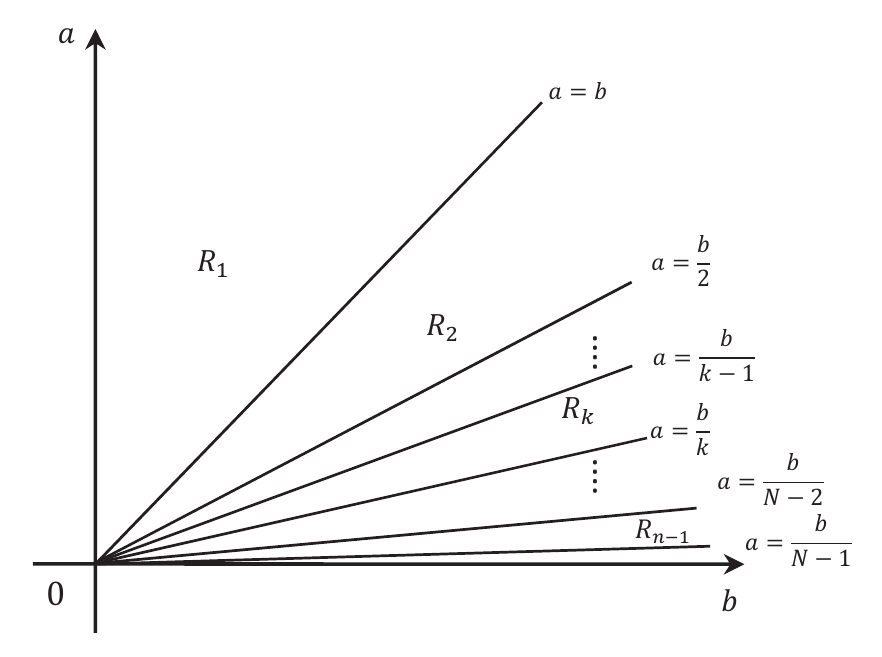}
	\caption{Feasible regions of the joint PDF of $\Am$ and $\Bm$.}
	\label{fig:R_range}
\end{figure}
From \eqref{CDF_SA}--\eqref{hn}, it is clear that the distribution of $\Bm$ conditioned on $\Am$ depends on the regions to which point $(a,b)$ belongs. We define regions $R_k$, $k=1,\ldots, N-1$, to be
\begin{align*}
	R_1 &= \left\{(a,b) \in \mathbb{R}^2_+: a \geq b \right\} \numberthis \label{R1},\\
	R_k &= \left\{(a,b) \in \mathbb{R}^2_+: \frac{b}{k} \leq a \leq \frac{b}{k-1} \right\}, \hspace{0.1cm} k = 2, \ldots, N-1. \numberthis \label{Rk}
\end{align*}
Fig. \ref{fig:R_range} illustrates the regions $R_k$, each of which corresponds to a different form of $h_N\left(\frac{a}{s}\right)$ in \eqref{hn}. We can then obtain the PDF of $\Bm$ conditioned on $\Am$ in the form of
\begin{align*}
\pdf_{\Bm|\Am}(b|a) &= \frac{e^{-b}}{ (1-e^{-a})^{N-1}(N-2)!} \\
 &\mathrel{\phantom{=}} \times \sum_{t=1}^k {N-1 \choose t-1}  (-1)^{t+1} [b-(t-1)a]^{N-2},\\ 
 &\mathrel{\phantom{=}} (a,b) \in R_k, \hspace{0.1cm} k=1,\ldots, N-1, \numberthis \label{fBgA}
\end{align*}
with the constraint $0 \leq b \leq (N-1)a$. From \eqref{fAm} and \eqref{fBgA}, the joint PDF of $\Am$ and $\Bm$ is expressed as
\begin{align*}
	\pdf_{\Am,\Bm}(a,b) 
	&= \pdf_{\Bm|\Am}(b|a) \pdf_{A_{max}}(a) \\
	&= \frac{N}{(N-2)!}e^{-(a+b)} \sum_{t=1}^k {N-1 \choose t-1} \\ 
	&\mathrel{\phantom{=}}\times (-1)^{t+1} [b-(t-1)a]^{N-2},\\
	&\mathrel{\phantom{=}} \hspace{3cm}(a,b) \in R_k. \numberthis \label{JoindPdfAmBm}
\end{align*}
For simplicity, we denote
\begin{align*}
	\pdf_k(a,b) = \pdf_{\Am,\Bm}(a,b),\hspace{0.1cm} (a,b) \in R_k.
\end{align*}
We now evaluate the integral in \eqref{I} by considering two cases:

\subsubsection{Case 1: $T\geq 1$}~
\begin{figure}[H]
	\centering
	\includegraphics[scale=1.2]{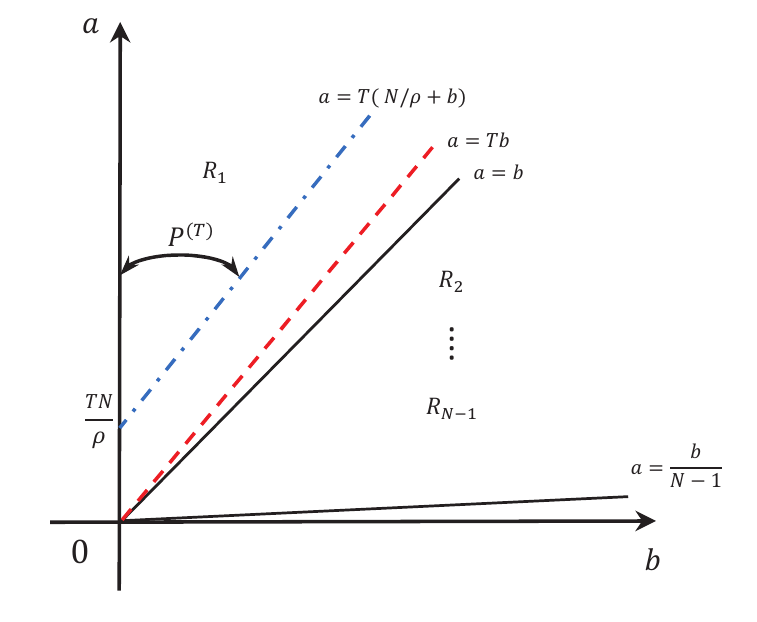}
	\caption{Feasible regions for determining  $\Pdl$ for $T \geq 1$.}
	\label{fig:R1}
\end{figure}
From \eqref{I} and Fig. \ref{fig:R1}, the downlink coverage probability in this case can be expressed as
\begin{align*}
	\Pdl = \int_{0}^{\infty} \int_{T\left(\frac{N}{\rho}+b\right)}^{\infty} f_1(a,b) dadb.
\end{align*}
In $R_1$, the joint distribution of $\Am$ and $\Bm$ in \eqref{JoindPdfAmBm} can be rewritten as
\begin{align}
\label{f1}
f_1(a,b) = \frac{N}{(N-2)!} e^{-(a+b)} b^{N-2}.
\end{align}
Hence, $\Pdl$ is expressed as
\begin{align*}
\Pdl 
&= \frac{N}{(N-2)!} \int_{0}^{\infty} \int_{T\left(\frac{N}{\rho}+b\right)}^{\infty} e^{-(a+b)} b^{N-2} dadb\\
&= \frac{N}{N-1}e^{-TN/\rho} \int_{0}^{\infty} e^{-(T+1)b} b^{N-2} db.
\end{align*}
By applying the partial integration, we obtain
\begin{align}
\label{par_int}
\int_{0}^{\infty} x^N e^{-\lambda x} = \frac{N!}{\lambda^{N+1}}, \hspace{0.1cm} \lambda > 0,
\end{align}
and hence, $\Pdl$ for $T \geq 1$ can be formulated as
\begin{align}
\label{Pdl_Tg1}
\Pdl = \frac{N}{(T+1)^{N-1}} e^{-TN/\rho}.
\end{align}
\subsubsection{Case 2: $0 < T < 1$}~
\begin{figure}[H]
	\centering
	\includegraphics[scale=1.0]{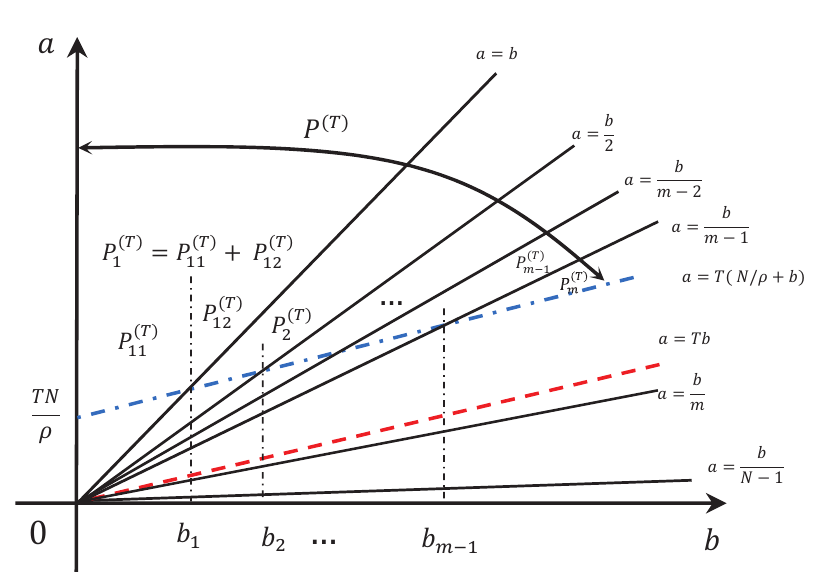}
	\caption{Feasible regions for determining  $\Pdl$ for $0 < T < 1$.}
	\label{fig:R2}
\end{figure}
From \eqref{I} and Fig. \ref{fig:R2}, the downlink coverage probability in this case can be expressed as
\begin{align}
	\label{Pdl_sum}
	\Pdl &= \Pdl_1 + \Pdl_2 + \ldots + \Pdl_{m-1} + \Pdl_m \nonumber\\
	&= \Pdl_1 + \sum_{i=2}^{m-1} \Pdl_i + \Pdl_m,
\end{align}
where 
\begin{align*}
	\Pdl_1 &= \underbrace{\int_0^{b_1} \int_{T\left(\frac{N}{\rho}+b\right)}^{\infty} f_1(a,b) dadb}_{:=\Pdl_{11}} + \underbrace{\int_{b_1}^{\infty} \int_{b}^{\infty} f_1(a,b) dadb}_{:=\Pdl_{12}}, \numberthis \label{Pdl1}\\
	\Pdl_k &= \underbrace{\int_{b_{k-1}}^{b_k} \int_{T\left(\frac{N}{\rho}+b\right)}^{\frac{b}{k-1}} f_k(a,b) dadb}_{:=\Pdl_{k1}} \\
	&\mathrel{\phantom{=}} + \underbrace{\int_{b_k}^{\infty} \int_{\frac{b}{k}}^{\frac{b}{k-1}} f_k(a,b) dadb}_{:=\Pdl_{k2}},\hspace{0.1cm} k = 2, \ldots, m-1 \numberthis \label{Pdlk}, \\
	\Pdl_m &= \int_{b_{m-1}}^{\infty} \int_{T\left(\frac{N}{\rho}+b\right)}^{\frac{b}{m-1}} f_m(a,b) dadb, \dis m = \left\lceil \frac{1}{T} \right\rceil, \numberthis \label{Pdlm}
\end{align*}
and $b_k$, $k=1,\ldots, N-1$, is the intersection point between the two lines $a = T\left(\frac{N}{\rho} + b\right)$ and $a = \frac{b}{k}$, as given by \eqref{bn_point}.
Inserting \eqref{f1} into \eqref{Pdl1}, we obtain
\begin{align*}
	\Pdl_1 
	&= \int_0^{b_1} \int_{T\left(\frac{N}{\rho}+b\right)}^{\infty} \frac{N}{(N-2)!} e^{-(a+b)} b^{N-2} \\
	&\mathrel{\phantom{=}} \hspace{2cm} + \int_{b_1}^{\infty} \int_{b}^{\infty}\frac{N}{(N-2)!} e^{-(a+b)} b^{N-2} dadb \\
	&= \frac{N}{(N-2)!} \left[e^{-\frac{TN}{\rho}} \underbrace{\int_{0}^{b_1}e^{-(T+1)b} b^{N-2} db}_{:=C_1} \right.\\
	&\mathrel{\phantom{=}} \left. \hspace{3.2cm} + \underbrace{\int_{b_1}^{\infty} e^{-2b} b^{N-2} db}_{:=C_2}\right]. \numberthis \label{Pdl1_proved}
\end{align*}
Using \eqref{par_int} and the lower incomplete Gamma function for an integer $n$, which is
\begin{align}
\label{gamma_f}
	\int_{\alpha}^{\infty} e^{-x}x^n dx = \Gamma(n+1,\alpha) = n! e^{-\alpha} \sum_{l=0}^n \frac{\alpha^l}{l!},
\end{align}
$C_1$ and $C_2$ can be expressed as
\begin{align*}
	C_1 
	&= \int_{0}^{b_1}e^{-(T+1)b} b^{N-2} db \\
	&= \int_{0}^{\infty}e^{-(T+1)b} b^{N-2} db - \int_{b_1}^{\infty}e^{-(T+1)b} b^{N-2} db\\
	&= \frac{(N-2)!}{(T+1)^{N-1}} \left(1-e^{-(T+1)b_1} \sum_{l=0}^{N-2} \frac{(T+1)^l b_1^l }{l!}\right), \\
	C_2 
	&= \int_{b_1}^{\infty}e^{-2b} b^{N-2} db = \frac{(N-2)!}{2^{N-1}} e^{-2b_1} \sum_{l=0}^{N-2} \frac{2^l b_1^l }{l!},
\end{align*}
which are the expressions in \eqref{C_1} and \eqref{C_2} in Appendix \ref{apd:CDEFGH}. Combining \eqref{C_1}, \eqref{C_2}, and \eqref{Pdl1_proved}, we obtain
\begin{align}
	\label{Pdl1_proved_cf}
\Pdl_1 	&= \frac{N}{(N-2)!} \left(e^{-\frac{TN}{\rho}} C_1 + C_2\right),
\end{align}
which is given by \eqref{Pdl_1} in Theorem \ref{theor:Pdl_closeform}.

We next evaluate $\Pdl_k$ in \eqref{Pdlk} by separately considering $\Pdl_{k1}$ and $\Pdl_{k2}$. Inserting \eqref{JoindPdfAmBm} into \eqref{Pdlk} yields
\begin{align}
	\label{Pdlk1}
	\Pdl_{k1} &= \frac{N}{(N-2)!} \sum_{t=1}^{k} {N-1 \choose t-1} (-1)^{t+1} \nonumber\\
	&\mathrel{\phantom{=}}\times \underbrace{\int_{b_{k-1}}^{b_k} \int_{T\left(\frac{N}{\rho}+b\right)}^{\frac{b}{k-1}} e^{-(a+b)} \left[b-(t-1)a\right]^{N-2} dadb}_{:=I}.
\end{align}
Using the binomial expansion, we have $[b-(t-1)a]^{N-2} = \sum_{i=0}^{N-2} {N-2 \choose i} b^{N-i-2} (1-t)^i a^i$. Hence, $I$ in \eqref{Pdlk1} becomes
\begin{align}
\label{Int_I}
	I &= \sum_{i=0}^{N-2} {N-2 \choose i} (1-t)^i \nonumber\\
	&\mathrel{\phantom{=}} \times \underbrace{\int_{b_{k-1}}^{b_k}  \underbrace{\int_{T\left(\frac{N}{\rho}+b\right)}^{\frac{b}{k-1}} e^{-a} a^i da}_{:=I_{ia}}  e^{-b} b^{N-i-2} db}_{:=I_i}.
\end{align}
Exploiting \eqref{gamma_f}, we obtain
\begin{align*}
	I_{ia} 
	&= \int_{T\left(\frac{N}{\rho}+b\right)}^{\infty} e^{-a} a^i da - \int_{\frac{b}{k-1}}^{\infty} e^{-a} a^i da \\
	&= i! \sum_{l=0}^{i} \left(e^{-T\left(\frac{N}{\rho}+b\right)} \frac{T^l\left(\frac{N}{\rho}+b\right)^l}{l!} - e^{-\frac{b}{k-1}} \frac{b^l}{(k-1)^l l!}\right).
\end{align*}
Hence, 
\begin{align}
\label{I1}
	I_i &= i! \sum_{l=0}^{i} \left(e^{-\frac{TN}{\rho}} \frac{T^l}{l!} \underbrace{\int_{b_{k-1}}^{b_k} e^{-(T+1)b} \left(\frac{N}{\rho}+b\right)^l b^{N-i-2} db}_{:=D_1} \nonumber \right. \\
	&\mathrel{\phantom{=}} \left. \hspace{1cm} - \frac{1}{(k-1)^l l!} \underbrace{\int_{b_{k-1}}^{b_k} e^{-\frac{k}{k-1}b} b^{N-i+l-2} db}_{:=D_2}\right).
\end{align}
Through steps similar to those for the derivations of $C_1$ and $C_2$, we can obtain the expressions for $D_1$ and $D_2$ as given in \eqref{D_1} and \eqref{D_2}. From \eqref{Pdlk1}--\eqref{I1}, and $\xi_p(\cdot)$ in \eqref{xi_f}, $\Pdl_{k1}$ can be rewritten as
\begin{align}
\label{Pdlk1_proved}
	\Pdl_{k1} = \xi_k \left(e^{-\frac{TN}{\rho}} \frac{T^l}{l!} D_1 - \frac{1}{(k-1)^l l!} D_2\right).
\end{align}

In a similar manner, the expressions of $\Pdl_{k2}$ and $\Pdl_m$ can also be derived as
\begin{align*}
\Pdl_{k2}
&= \xi_k \left(e^{-\frac{TN}{\rho}} \frac{T^l}{l!} E_1 - \frac{1}{(k-1)^l l!} E_2\right), \numberthis \label{Pdlk2_proved}\\
\Pdl_m 	&= \xi_m \left(e^{-\frac{TN}{\rho}} \frac{T^l}{l!} F_1 - \frac{1}{(m-1)^l l!} F_2		\right) \numberthis \label{Pdlm_proved},
\end{align*}
where $E_1$, $E_2$, $F_1$, and $F_2$ are given in \eqref{E_1}--\eqref{F_2} in Appendix \ref{apd:CDEFGH}. Finally, Theorem \ref{theor:Pdl_closeform} is proved by combining \eqref{Pdl_sum},  \eqref{Pdl1_proved_cf}, and \eqref{Pdlk1_proved}--\eqref{Pdlm_proved}.

\section{Prove of Remark \ref{rm:varying_property_T_greater_than_1}}
\label{apd:proof_rmk2}
By Theorem \ref{theor:Pdl_closeform}, when $T \geq 1$, the downlink coverage probability is written as
\begin{align}
\label{PdlTgreaterthan1}
\Pdl = \frac{N}{\left(T+1\right)^{N-1}} e^{-\frac{TN}{\rho}}.
\end{align}
We observe that
\begin{align}
\label{cond_1}
\frac{N}{\left(T+1\right)^{N-1}} \leq \frac{N}{2^{N-1}} \leq 1,\hspace{0.1cm} T \geq 1, N \geq 1,
\end{align}
where the equalities simultaneously occur for $N=1$. Furthermore, we have
\begin{align}
\label{cond_2}
e^{-\frac{TN}{\rho}} \leq e^{-\frac{T}{\rho}},\hspace{0.1cm} T \geq 1, N \geq 1,
\end{align}
where the equality also occurs for $N=1$. From \eqref{cond_1} and \eqref{cond_2}, we obtain
\begin{align}
\label{rs1}
\Pdl = \frac{N}{\left(T+1\right)^{N-1}} e^{-\frac{TN}{\rho}} \leq e^{-\frac{T}{\rho}},\hspace{0.1cm} T \geq 1, N \geq 1,
\end{align}
where the equality occurs for $N=1$.
Therefore, we conclude that when $T \geq 1$, the optimal number of precoding vectors at which the downlink coverage probability is maximized is $N^*=1$.

We now prove that $\Pdl$ with $T \geq 1$ is a decreasing function of $N$. We observe that
\begin{align*}
\frac{\partial \Pdl}{\partial N}  
&= \frac{\partial}{\partial N} \left(\frac{N}{\left(T+1\right)^{N-1}} e^{-\frac{TN}{\rho}} \right)\\
&= -\frac{e^{-\frac{TN}{\rho}}}{(T+1)^{N-1}} \left(\frac{TN}{\rho} + N \log(T+1) - 1\right) \\
&< 0,\hspace{0.1cm} T \geq 1, N \geq 2. \numberthis \label{derivation_sign}
\end{align*}
Hence, $\Pdl$ is maximum at $N=1$ and is a decreasing function of $N$ on the range $[2;\infty)$.

The decreasing rate of $\Pdl$ with respect to $N$ can be formulated as
\begin{align*}
\zeta = \left|\frac{\partial \Pdl}{\partial N} \right| = \frac{e^{-\frac{TN}{\rho}}}{(T+1)^{N-1}} \left(\frac{TN}{\rho} + N \log(T+1) - 1\right), \hspace{0.1cm} \\T \geq 1.
\end{align*}
In addition, the derivative of $\zeta$ with respect to $N$ is expressed as
\begin{align*}
\frac{\partial \zeta}{\partial N}  
&= \frac{\partial}{\partial N} \left(\frac{e^{-\frac{TN}{\rho}}}{(T+1)^{N-1}} \left(\frac{TN}{\rho} + N \log(T+1) - 1\right) \right)\\
&= -\frac{e^{-\frac{TN}{\rho}}}{(T+1)^{N-1}} \left(\frac{T}{\rho} +  \log(T+1)\right) \\
&\mathrel{\phantom{=}} \hspace{1.9cm}\times \left(\frac{TN}{\rho} + N \log(T+1) - 2\right) \numberthis \label{derivation_veta},
\end{align*}
which has a single zero at
\begin{align}
N_0 = \frac{2}{T/\rho + \log(T+1)}
\end{align}
and is negative on $[N_0; \infty)$.
For $T \geq 1$, we have
\begin{align}
N_0 \leq \frac{2}{1/\rho + \log(2)} < \frac{2}{\log(2)} < 3.
\end{align}
Therefore, we can conclude that for the range $[3; \infty)$, $\zeta$ is a decreasing function of $N$. 

We now prove that $\Pdl$ decreases on $[2;3]$ faster than on $[3;4]$. From \eqref{PdlTgreaterthan1}, the values of $\Pdl$ at $N=2$, $N=3$, and $N=4$ are
\begin{align*}
\Pdl_2 &= \frac{2}{T+1} e^{-\frac{T}{\rho}},\\
\Pdl_3 &= \frac{3}{\left(T+1\right)^{2}} e^{-\frac{2T}{\rho}},\\
\Pdl_4 &= \frac{4}{\left(T+1\right)^{3}} e^{-\frac{3T}{\rho}},
\end{align*}
respectively. From the decreasing property of $\Pdl$, we have
$
\Pdl_2 > \Pdl_3 > \Pdl_4.
$
Therefore, the decreasing rates of $\Pdl$ on $[2;3]$ and $[3;4]$ are determined to be
\begin{align*}
\left|	\frac{\Delta \Pdl_{2,3}}{\Delta N} \right| &= \Pdl_2 - \Pdl_3 =  \underbrace{\frac{e^{-\frac{2T}{\rho}}}{T+1}}_{:= \alpha_1} \underbrace{\left(2 - \frac{3}{T+1}e^{-\frac{T}{\rho}} \right)}_{:= \beta_1},\numberthis \label{delta_23}\\
\left|	\frac{\Delta \Pdl_{3,4}}{\Delta N} \right | &= \Pdl_3 - \Pdl_4 =  \underbrace{\frac{e^{-\frac{3T}{\rho}}}{(T+1)^2}}_{:= \alpha_2} \underbrace{\left(3 - \frac{4}{T+1}e^{-\frac{T}{\rho}} \right)}_{:= \beta_2},\numberthis \label{delta_34}
\end{align*}
respectively. We observe that
\begin{align*}
\frac{\alpha_1}{\alpha_2} &= e^{\frac{T}{\rho}} (T+1) > 2, \numberthis \label{frac_alpha} \hspace{0.1cm} T \geq 1,\\
\frac{\beta_2}{\beta_1} &= 2 - \underbrace{\frac{1 - \frac{2}{T+1} e^{-\frac{T}{\rho}}}{2 - \frac{3}{T+1} e^{-\frac{T}{\rho}}}}_{:= \kappa}, \hspace{0.1cm} T \geq 1.
\end{align*}
Furthermore, we have
\begin{align*}
\frac{2}{T+1} e^{-\frac{T}{\rho}} &< 1, \hspace{0.1cm} T \geq 1,\\
\frac{3}{T+1} e^{-\frac{T}{\rho}} &< \frac{3}{2}, \hspace{0.1cm} T \geq 1,
\end{align*}
which lead to $\kappa > 0$, and hence
\begin{align}
\label{frac_beta}
\frac{\beta_2}{\beta_1} < 2, \hspace{0.1cm} T \geq 1.
\end{align}
From \eqref{delta_23} and \eqref{delta_34}, $\alpha_1$, $\alpha_1$, $\beta_1$, and $\beta_2$ are positive. Therefore, from \eqref{frac_alpha} and \eqref{frac_beta}, we have $\alpha_1 \beta_1 > \alpha_2 \beta_2$, which means that
$
\frac{\Delta \Pdl_{2,3}}{\Delta N} > \frac{\Delta \Pdl_{3,4}}{\Delta N}.
$
In other words, $\Pdl$ decreases faster on $[2;3]$ than on $[3;4]$, which in conjunction with the decreasing property of $\zeta = \left|\frac{\partial \Pdl}{\partial N} \right|$ on $[3; \infty)$ leads to the conclusion that $\Pdl$ decreases more slowly with $N$ on $[2, \infty)$. Thus, the proof of Remark \ref{rm:varying_property_T_greater_than_1} is complete.
%
%

\bibliographystyle{IEEEtran}
\bibliography{IEEEabrv,Bibliography}

%
%
%

\end{document}